\documentclass{article}

\usepackage{microtype}
\usepackage{graphicx}
\usepackage{subfigure}
\usepackage{color}

\usepackage{caption,subfigure,graphicx,epstopdf,multirow,multicol,booktabs,verbatim,wrapfig,bm}
\usepackage{booktabs} 

\usepackage{hyperref}
\usepackage{longtable}
\usepackage{booktabs}
\usepackage{algorithmic}
\usepackage{graphicx}
\usepackage{subfigure}
\usepackage{textcomp}
\usepackage{booktabs}
\usepackage{amsmath,amscd,amsbsy,amssymb,amsfonts,latexsym,url,bm,amsthm}
\usepackage{multirow,multicol,array}

\usepackage{threeparttable}

\usepackage[accepted]{icml2021}

\icmltitlerunning{An Inductive Model-based Collaborative Filtering Approach}

\begin{document}

\twocolumn[
\icmltitle{Towards Open-World Recommendation: \texorpdfstring{\\}{a}
An Inductive Model-based Collaborative Filtering Approach}



\icmlsetsymbol{equal}{*}

\begin{icmlauthorlist}
\icmlauthor{Qitian Wu}{sjtu,moe}
\icmlauthor{Hengrui Zhang}{sjtu}
\icmlauthor{Xiaofeng Gao}{sjtu,moe}
\icmlauthor{Junchi Yan}{sjtu,moe}
\icmlauthor{Hongyuan Zha}{cuhk-sz}
\end{icmlauthorlist}

\icmlaffiliation{sjtu}{Department of Computer Science and Engineering, Shanghai Jiao Tong University.}
\icmlaffiliation{moe}{MoE Key Lab of Artificial Intelligence, AI Institute, Shanghai Jiao Tong University.}
\icmlaffiliation{cuhk-sz}{School of Data Science, Shenzhen Institute of Artificial Intelligence and Robotics for Society, The Chinese University of Hong Kong, Shenzhen}

\icmlcorrespondingauthor{Xiaofeng Gao}{gao-xf@cs.sjtu.edu.cn}


\vskip 0.3in
]

\printAffiliationsAndNotice{} 

\begin{abstract}
  Recommendation models can effectively estimate underlying user interests and predict one's future behaviors by factorizing an observed user-item rating matrix into products of two sets of latent factors. However, the user-specific embedding factors can only be learned in a transductive way, making it difficult to handle new users on-the-fly.
  In this paper, we propose an inductive collaborative filtering framework that contains two representation models. The first model follows conventional matrix factorization which factorizes a group of key users' rating matrix to obtain \emph{meta latents}. The second model resorts to attention-based structure learning that estimates hidden relations from query to key users and learns to leverage meta latents to inductively compute embeddings for query users via neural message passing. Our model enables inductive representation learning for users and meanwhile guarantees equivalent representation capacity as matrix factorization. Experiments demonstrate that our model achieves promising results for recommendation on few-shot users with limited training ratings and new unseen users which are commonly encountered in open-world recommender systems. The codes are available at \url{https://github.com/qitianwu/IDCF}. 
\end{abstract}

\section{Introduction}\label{sec-intro}
As information explosion has become one major factor affecting human life, recommender systems, which can filter useful information and contents of user's potential interests, play an increasingly indispensable role. Recommendation problems can be generally formalized as matrix completion (MC) where one has a user-item rating matrix whose entries, which stand for interactions of users with items (ratings or click behaviors), are partially observed. The goal of MC is to predict missing entries (unobserved or future potential interactions) in the matrix based on the observed ones. 

Modern recommender systems need to meet two important requirements  for practical utility. First of all, \emph{recommendation models should have enough expressiveness to capture diverse user interests and preferences so that the systems can accomplish personalized recommendation}. 
Existing methods based on collaborative filtering (CF)\footnote{In recent literature, collaborative filtering (CF) approaches often refer to model-based CF, i.e. matrix factorization, while its memory-based counterpart, as an heuristic approach, adopts similarity methods like KNN for recommendation.} or, interchangeably, matrix factorization (MF) have shown great power in this problem by factorizing the rating matrix into two classes of latent factors (i.e., embeddings) for users and items respectively, and further leverage dot-product of two factors to predict potential ratings \cite{Imp-early,MFrec,BPR,MFGenError,MFAutoReg}. Equivalently, for each user, the methods consider a one-hot user encoding as input, and assume a user-specific embedding function mapping user index to a latent factor. Such learnable latent factors can represent user's preferences in a low-dimensional space. 
Recent works extend MF with complex architectures, like multi-layer perceptrons \cite{NNMF}, recurrent units \cite{sRGCNN}, graph neural networks \cite{GCMC,dancer-wu}, etc., achieving state-of-the-art results on both benchmark datasets and commercial systems \cite{youtube,PinSage,quantized-emb}.

\begin{figure*}[t]
	\centering
	\includegraphics[width=0.98\textwidth]{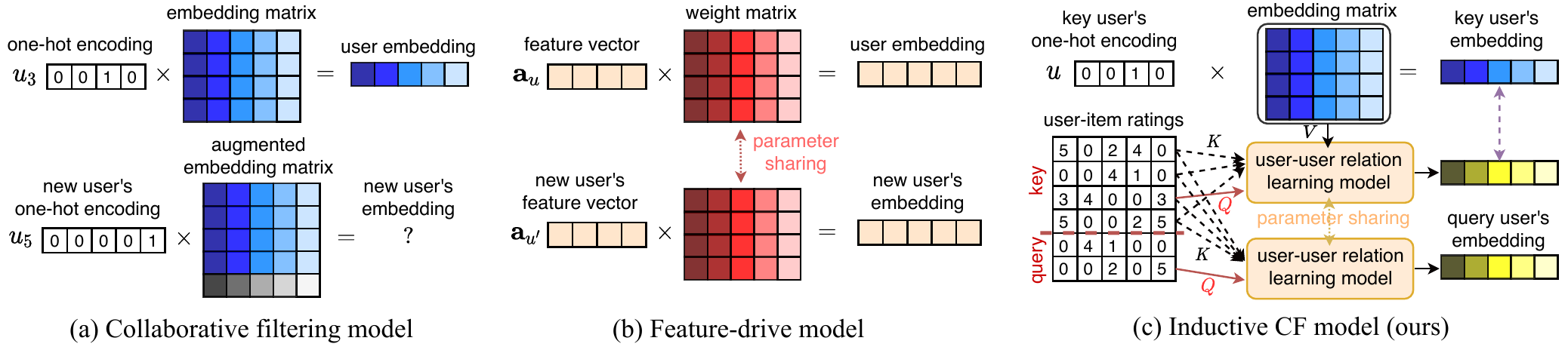}
	\vspace{-10pt}
	\caption{(a) Conventional (model-based) collaborative filtering model assumes user-specific embeddings for user representation, which is parameterized in a user-specific manner and limits the model's capability to handle new unseen users. (b) Feature-driven models manage to deal with new users and achieves inductive learning via modeling a user-sharing mapping from user features to representations, but would lack enough expressiveness for diverse user interests. (c) Our proposed inductive CF model that absorbs the advantage of both of the worlds, achieving inductive representation learning for users without compromising representation capacity.}
	\label{fig-intro}
		\vspace{-10pt}
\end{figure*}

The second requirement stems from a key observation from real-world scenarios: \emph{recommender systems often interact with a dynamic open world where new users, who are not exposed to models during training, may appear in test stage}. This requires that models trained on a set of users manage to adapt to unseen users. However, the above-mentioned CF models would fail in this situation since the embedding factors are parameterized for specific users and need to be learned collaboratively with all other users in transductive setting (see Fig.~\ref{fig-intro} for illustration). Brute-force ways include: 1) retrain a model with an augmented rating matrix; 2) consider incremental learning \cite{Imp-early} for new users' embeddings. The former requires much extra time cost, which would be unacceptable for online systems, while the latter is prone for over-fitting and disables on-the-fly inference. There are quite a few studies that propose inductive matrix completion models using user features \cite{IMC, Feature1, Feature2, PinSage, NIMC}. Their different thinking paradigm, as shown in Fig.~\ref{fig-intro}(b), is to target a user-sharing mapping from user features to user representations, instead of from one-hot user indices. Since the feature space is shared among users, such methods are able to adapt a trained model to unseen users. Nevertheless, feature-driven models may suffer from limited expressiveness with low-quality features that have weak correlation with labels. For example, users with the same age and occupation (commonly used features) may have distinct ratings on movies. Unfortunately, high-quality features that can unveil user interests for personalized recommendation are often hard to collect due to increasingly concerned privacy issues.


A following question arises: \emph{Can we build a recommendation model that guarantees enough expressiveness for personalized preferences and enables inductive learning?} Such question still remains unexplored so far. In fact,  simultaneously meeting the two requirements is a non-trivial challenge when high-quality user features are unavailable. First, to achieve either of them, one often needs to compromise on the other. The user-specific embedding vectors, which assume independent parametrization for different users, can give sufficient capacity for learning distinct user preferences from historical rating patterns \cite{word2vec}. To make inductive learning possible, one needs to construct a shared input feature space among users out of the rating matrix, as an alternative to one-hot user encodings. However, the new constructed features have relatively insufficient expressive power. Second, the computation based on new feature space often brings extra costs for time and space, which limits model's scalability on large-scale datasets.


In this paper, we propose an InDuctive Collaborative Filtering model (IDCF) as a general CF framework that achieves inductive learning for user representations and meanwhile guarantees enough expressiveness and scalability, as shown in Fig.~\ref{fig-intro}. Our approach involves two representation models. A conventional matrix factorization model that factorizes a group of key users' rating matrix to obtain their user-specific embeddings, which we call \emph{meta latents}. On top of that, we further design a relation learning model, specified as multi-head attention mechanism, that learns hidden graphs between key and query users w.r.t. their historical rating behaviors. The uncovered relation graphs enable neural message passing among users in latent space and inductive computation of user-specific representations for query users. 

Furthermore, we develop two training strategies for practical cases in what we frame as open-world recommendation: inductive learning for interpolation and inductive learning for extrapolation, respectively. In the first case, query users are disjoint from key users in training and the model is expected to provide decent performance on few-shot query users with limited training ratings. In the second case, query users are the same as key users in training and the trained model aims to tackle zero-shot test users that are unseen before. We show that our inductive model guarantees equivalent capacity as matrix factorization and provides superior expressiveness compared to other inductive models (feature-driven, item-based and graph-based). Empirically, we conduct experiments on five datasets for recommendation (with both explicit and implicit feedbacks). The comprehensive evaluation demonstrates that IDCF 1) consistently outperform various inductive models by a large margin on recommendation for few-shot users , and 2) can achieve superior results on recommendation for new unseen users (with few historical ratings not used in training). Moreover, compared with transductive models, IDCF provides very close reconstruction error and can even outperform them when training ratings becomes scarce. The contributions of this paper are summarized as follows.

1) We propose a new inductive collaborative filtering framework that can inductively compute user representations for new users based on a set of pretrained meta latents, which is suitable for open-world recommender systems. The new approach can serve as a brand new learning paradigm for inductive representation learning.

2) We show that a general version of our model can minimize reconstruction loss to the same level as vanilla matrix factorization model under a mild condition. Empirically, IDCF gives very close RMSE to transductive CF models.

3) IDCF achieves very competitive and superior RMSE/NDCG/AUC results on few-shot and new unseen users compared with various inductive models on explicit feedback and implicit feedback data. 

As a general model-agnostic framework, IDCF can flexibly incorporate with various off-the-shelf MF models (e.g. MLP-based, GNN-based, RNN-based, attention-based, etc.) as backbones as well as combine with user profile features for hybrid recommendation.

\section{Background}
 Consider a general matrix completion (MC) problem which deals with a user-item rating matrix $R=\{r_{ui}\}_{M\times N}$ where $M$ and $N$ is the number of users and items, respectively. For explicit feedback, $r_{ui}$ records rating 
value of user $u$ on item $i$. For implicit feedback, $r_{ui}$ is a binary entry for whether user $u$ rated (or clicked on, reviewed, liked, purchased, etc.) item $i$ or not. The recommendation problem can be generally formalized as: given partially observed entries in $R$, one needs to estimate the missing values in the matrix.

Existing recommendation models are mostly 
based on collaborative filtering (CF) or, interchangeably, matrix factorization (MF) where user $u$ (resp. item $i$) corresponds to a $d$-dimensional latent factor (i.e., embedding) $\mathbf p_u$ (resp. $\mathbf q_i$). 
Then one has a prediction model $\hat r_{ui}=f(\mathbf p_u,\mathbf q_i)$ where $f$ can be basically specified as simple dot product or some complex architectures, like neural networks, graph neural networks, etc. One advantage of CF models is that the user-specific embedding $\mathbf p_u$ (as learnable parameters) can provide enough expressive power for learning diverse personal preferences from user historical behaviors and decent generalization ability through collaborative learning with all the users and items. However, such user-specific parametrization limits the model in transductive learning. In practical situations, one cannot have information for all the users that may appear in the future when collecting training data. When it comes to new users in test stage, the model has to be retrained and cannot deliver on-the-fly inference. 

\begin{figure}[t!]
	\centering
	\includegraphics[width=0.48\textwidth]{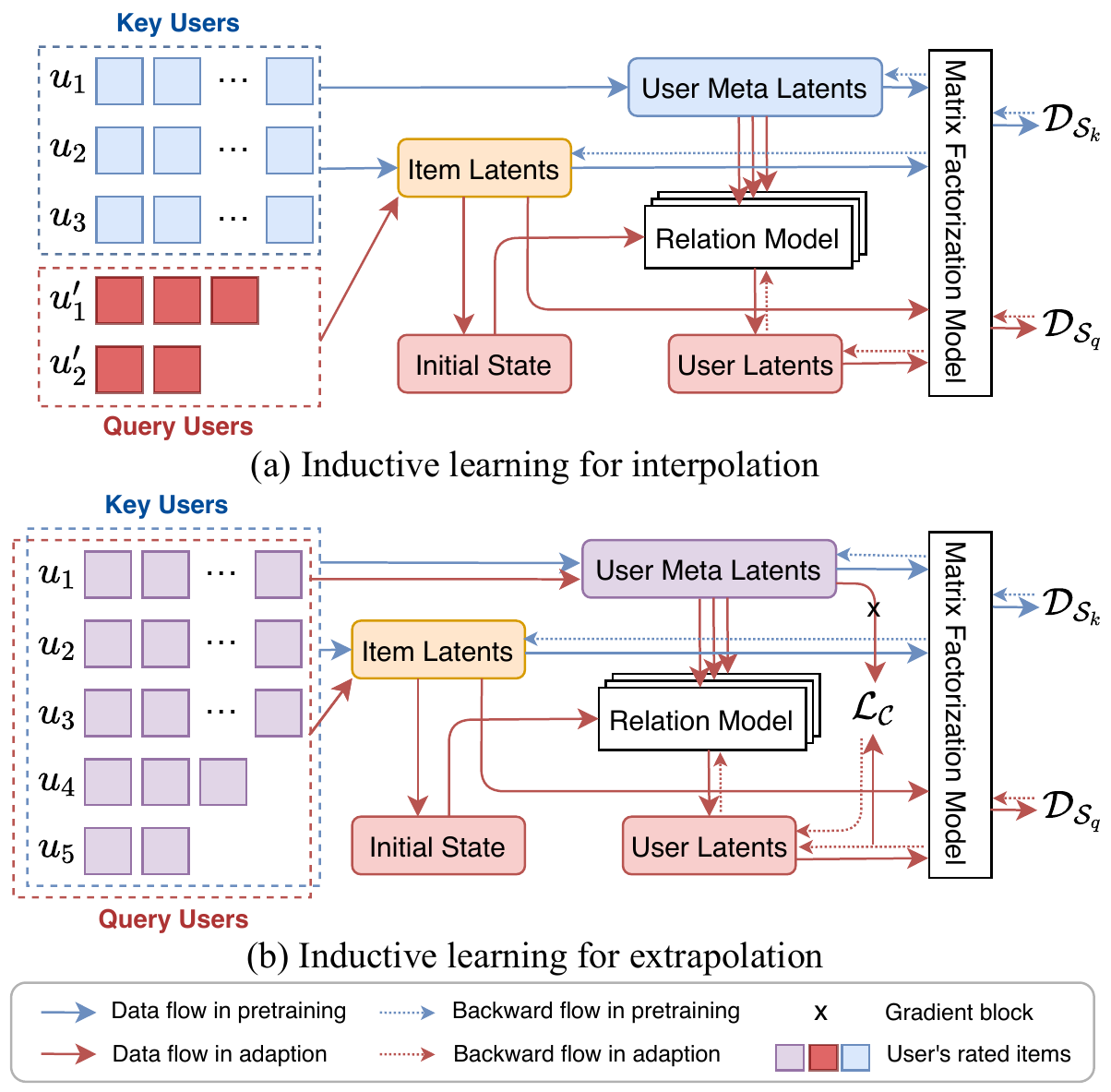}
	\vspace{-20pt}
	\caption{Framework of inductive collaborative filtering for open-world recommendation. We consider two scenarios, inductive learning for interpolation (query users are different from key users) and extrapolation (query users are the same as key users), which aims to handle few-shot query users and zero-shot new users in test stage. In both cases, the learning procedures contain pretraining and adaption. The pretraining learns initial user meta latents via matrix factorization over key users' ratings. The adaption optimizes a relation model, which estimates hidden relations from key to query users and inductively computes embeddings for query users. In particular, in extrapolation case, we introduce a self-supervised contrastive loss that enforces similarity between meta latents and inductively computed embeddings for the same users.}
	\label{fig-model}
		\vspace{-10pt}
\end{figure}

\section{Methodology}\label{sec-model}
We propose the InDuctive Collaborative Filtering (IDCF) model. 
Our high-level methodology stems from a key observation: there exist a (or multiple) latent relational graph among users that represents preference proximity and behavioral interactions. For instance, social networks and following networks in social media can be seen as realizations of such relational graphs, but in most cases, the graph structures are unobserved and implicitly affect user's behaviors. If we can identify the graph structures, we can leverage the idea of message passing \cite{GNN-early1,graphsage,chemi-messpass,dancer-wu}, propagating learned embeddings from one group of users to others, especially, in an inductive manner. 

We formulate our model through two sets of users: 1) \emph{key users} (denoted by $\mathcal U_k$), for which we learn their embeddings by matrix factorization and use them as \emph{meta latents}; 2) \emph{query users} (denoted by $\mathcal U_q$), for which we consider neural message passing to inductively compute their embeddings. Assume $|\mathcal U_k|=M_k$ and $|\mathcal U_q|=M_q$. Correspondingly, we have two rating matrices: $R_k=\{r_{ui}\}_{M_k\times N}$ (given by $\mathcal U_k$) and $R_q=\{r_{u'i}\}_{M_q\times N}$ (given by $\mathcal U_q$). Based on this, we further consider two scenarios.

\textbf{Scenario I}: Inductive learning for interpolation. $\mathcal U_1 \cap \mathcal U_2 = \emptyset$, i.e., query users are disjoint from key users. In training, it learns to leverage meta latents of key users to compute representations for another group of query users in a supervised way. The model is expected to perform robustly on few-shot query users with limited training ratings.

\textbf{Scenario II}: Inductive learning for extrapolation. $\mathcal U_1 = \mathcal U_2$, i.e., key and query users are the same. In training, it learns to use meta latents given by key users to represent themselves in a self-supervised way. Then the trained model aims to deal with zero-shot test users (with limited observed ratings not used in training) that are unseen before.

Notice that in the above two cases, we assume no side information, such as user profile features (ages, occupation, etc,), social networks, item content features, etc., besides the observed user-item rating matrix. We frame the problem as open-world recommendation which requires the model to deal with few-shot and zero-shot users. We present our model framework for two settings in Fig.~\ref{fig-model} and go into the details in the following.

\subsection{Matrix Factorization Model}\label{sec-model1}
We first pretrain a (transductive) matrix factorization model for $\mathcal U_k$ using $R_k$, denoted as $\hat r_{ui}=f_\theta(\mathbf p_u,\mathbf q_i)$, where $\mathbf p_u\in \mathbb R^d$ denotes a user-specific embedding for user $u$ in $\mathcal U_k$, $\mathbf q_i\in \mathbb R^d$ denotes an item-specific embedding for item $i$ and $f_\theta$ can be simple dot-product or a network with parameter $\theta$.  
Section~\ref{sec-setup}
gives details for two specifications for $f_\theta$ using neural network and graph convolution network, as used in our implementation. Denote $\mathbf P_k=\{\mathbf p_u\}_{M_k\times d}$, $\mathbf Q=\{\mathbf q_i\}_{N\times d}$ and the objective becomes
\begin{equation}\label{eqn-obj1}
\min\limits_{\mathbf P_k, \mathbf Q, \theta} \mathcal D_{\mathcal S_k}(\hat R_k, R_k),
\end{equation}
where we define $\hat R_k=\{\hat r_{ui}\}_{M_k\times N}$, $\mathcal D_{\mathcal S_k}(\hat R_k, R_k)=\frac{1}{T_k}\sum_{(u,i)\in\mathcal S_k} l(r_{ui}, \hat r_{ui})$ and $\mathcal S_k\in ([M_k]\times[N])^{T_k}$ is a set with size $T_k$ containing indices of observed entries in $R_k$. Here $l(r_{ui}, \hat r_{ui})$ can be MSE loss for explicit user feedbacks or cross-entropy loss for implicit user feedbacks. 


We treat the pretrained factors $\mathbf P_k$ as \emph{meta latents}, to inductively compute user latents for query via a relation model.

\subsection{Inductive Relation Model}\label{sec-model2}
Assume $\mathbf C=\{c_{uu'}\}_{M_k\times M_q}$, where $c_{uu'}\in \mathbb R$ denotes weighted edge from user $u\in \mathcal U_k$ to user $u'\in \mathcal U_q$, and define $\mathbf c_{u'}=[c_{1u'}, c_{2u'}, \cdots c_{M_ku'}]^\top$ the $u'$-th column of $\mathbf C$. Then we express embedding of user $u'$ as $\mathbf {\tilde p}_{u'}= \mathbf c_{u'}^\top \mathbf P_k$, a weighted sum of embeddings of key users. The rating can be predicted by $\hat r_{u'i}=f_\theta(\mathbf {\tilde p}_{u'}, \mathbf q_i)$ and the problem of model optimization becomes:
\begin{equation}\label{eqn-obj2}
\min\limits_{\mathbf C, \mathbf Q} \mathcal D_{\mathcal S_q}(\hat R_q, R_q),
\end{equation}
where we define $\hat R_q=\{\hat r_{u'i}\}_{M_q\times N}$, $\mathcal D_{\mathcal S_q}(\hat R_q, R_q)=\frac{1}{T_q}\sum_{(u',i)\in\mathcal S_q} l(r_{u'i}, \hat r_{u'i})$ and $\mathcal S_q\in ([M_q]\times[N])^{T_q}$ is a set with size $T_q$ containing indices of observed entries in $R_q$.
The essence of above method is taking attentive pooling as message passing from key to query users. We first justify this idea by analyzing its capacity and then propose a parameterized model  enabling inductive learning.

\textbf{Theoretical Justification}
If we use dot-product for $f_\theta$ in the MF model, then we have $\hat r_{u'i}=\mathbf {\tilde p}_{u'}^\top \mathbf q_i$. We compare Eq.~(\ref{eqn-obj2}) with using matrix factorization over $R_q$:
\begin{equation}\label{eqn-obj3}
\min\limits_{\mathbf {\tilde P}_q, \mathbf Q} \mathcal D_{\mathcal S_q}(\hat R_q, R_q),
\end{equation}
where $\mathbf {\tilde P}_q=\{\mathbf {\tilde p}_{u'}\}_{M_q\times d}$ and have the following theorem.
\paragraph{Theorem 1.}
\emph{
	Assume Eq.~(\ref{eqn-obj3}) can achieve $\mathcal D_{\mathcal S_q}(\hat R_q, R_q)<\epsilon$ and the optimal $\mathbf P_k$ given by Eq.~(\ref{eqn-obj1}) satisfies column-full-rank, then there exists at least one solution for $\mathbf C$ in Eq.~(\ref{eqn-obj2}) such that $\mathcal D_{\mathcal S_q}(\hat R_q, R_q)<\epsilon$.
}

The only condition that $\mathbf P_k$ is column-full-rank can be trivially guaranteed since $d \ll N$. 
The theorem shows that the proposed 
model can minimize the reconstruction loss of MC to at least the same level as matrix factorization which gives sufficient capacity for learning personalized user preferences from historical rating patterns.

\textbf{Parametrization}
We have shown that using attentive pooling does not sacrifice model capacity than MF/CF models under a mild condition. However, directly optimizing over $\mathbf C$ is intractable due to its $O(M_kM_q)$ parameter space and $\mathbf c_{u'}$ is a user-specific high-dimensional vector which disallows inductive learning. Hence, we parametrize 
$\mathbf C$ with an attention network, reducing parameters and enabling it for inductive learning. Concretely, we estimate the adjacency score between user $u'$ and $u$ as
\begin{equation}\label{eqn-attn}
    c_{u'u}=\frac{ \mathbf e^\top[\mathbf W_q \mathbf d_{u'}\oplus \mathbf W_k\mathbf p_u]}{\sum_{u_0\in\mathcal U_1} \mathbf e^\top[\mathbf W_q\mathbf d_{u'}\oplus \mathbf W_k\mathbf p_{u_0}]},
\end{equation}
where $\mathbf e\in \mathbb R^{2d\times 1}$, $\mathbf W_q \in\mathbb R^{d\times d} $, $\mathbf W_k \in\mathbb R^{d\times d} $ are trainable parameters, $\oplus$ denotes concatenation and $\mathbf d_{u'}=\sum_{i\in \mathcal I_{u'}}\mathbf q_i$. Here $\mathcal I_{u'}=\{i|r_{u'i} > 0\}$ includes the historically rated items of user $u'$. The attention network captures first-order user proximity on behavioral level and also maintains second-order proximity that users with similar historical ratings on items would have similar relations to other users. Besides, if $\mathcal I_{u'}$ is empty (for extreme cold-start recommendation), we can randomly select a group of items from all the candidates. Also, if user's profile features are available, we can harness them as $\mathbf d_{u'}$. We provide details in Appendix~\ref{appendix-extension}. Yet, in the main body of our paper, we focus on learning from user-item rating matrix, i.e., the common setting for CF approaches.

The normalization in Eq.~(\ref{eqn-attn}) requires computation for all the key users, which limits scalability to large dataset. Therefore, we use sampling strategy to control the size of key users in relation graph for each query user and further consider multi-head attentions that independently sample different subsets of key users. The attention score given by the $l$-th head is
\begin{equation}\label{eqn-attn2}
    c_{u'u}^{(l)}=\frac{ (\mathbf e^{(l)})^\top[\mathbf W_q^{(l)} \mathbf d_{u'}\oplus \mathbf W_k^{(l)}\mathbf p_u]}{\sum_{u_0\in\mathcal U_k^{(l)}} (\mathbf e^{(l)})^\top[\mathbf W_q^{(l)}\mathbf d_{u'}\oplus \mathbf W_k^{(l)}\mathbf p_{u_0}]},
\end{equation}
where $\mathcal U_k^{(l)}$ denotes a subset of key users sampled from $\mathcal U_k$. Each attention head independently aggregates embeddings of different subsets of key users and the final inductive representation for user $u'$ can be given as
\begin{equation}\label{eqn-induc_embed}
    \mathbf {\tilde p}_{u'}=\mathbf W_o\left[ \bigoplus\limits_{l=1}^L \sum_{u\in\mathcal U_k^{(l)}}c_{u'u}^{(l)} \mathbf W_v^{(l)}\mathbf p_u\right],
\end{equation}
where $\mathbf W_o\in \mathbb R^{d\times Ld}$ and $\mathbf W_v^{(l)}\in \mathbb R^{d\times d}$. To keep the notation clean, we denote $\mathbf  {\tilde p}_{u'}=h_w(\mathbf d_{u'})$ and $w=\cup_{l=1}^L \{\mathbf e^{(l)}, \mathbf W_q^{(l)}, \mathbf W_k^{(l)}, \mathbf W_v^{(l)}\} \cup \{\mathbf W_o\}$. 

With fixed meta latents $\mathbf P_1$ and $\mathbf Q$, we can consider optimization for our inductive relation model
\begin{equation}
    \min_{w, \theta}\mathcal D_{\mathcal S_q}(\hat R_q, R_q).
\end{equation}
We found that using fixed $\mathbf Q$ here contributes to much better performance than optimizing it in the second stage. 

Next we analyze the generalization ability of inductive model on query users. Also, consider $f_\theta$ as dot-product operation and we assume $c_{u'u}\in \mathbb R^+$ to simplify the analysis. Now, we show that the generalization error $\mathcal D(\hat R_q, R_q)= \mathbb E_{(u',i)} [l(r_{u'i}, \hat r_{u'i})]$ on query users is bounded by the numbers of key users and observed ratings of query users.
\paragraph{Theorem 2.}
\emph{
	Assume 1) $\mathcal D$ is $L$-Lipschitz, 2) for $\forall \hat r_{u'i}\in \hat R_q$ we have $|\hat r_{u'i}|\leq B$, and 3) the L1-norm of $\mathbf c_{u'}$ is bounded by $H$. Then with probability at least $1-\delta$ over the random choice of $\mathcal S_q\in ([M_q]\times [N])^{T_q}$, it holds that for any $\hat R_q$, the gap between $\mathcal D(\hat R_q, R_q)$ and $\mathcal D_{\mathcal S_q}(\hat R_q, R_q)$ will be bounded by}
	\begin{equation}
     O\left(2LHB\sqrt{\frac{2M_q\ln M_k}{T_q}}+\sqrt{\frac{ln(1/\delta)}{T_q}}\right).
	\end{equation}

The theorem shows that the generalization error bound depends on the size of $\mathcal U_k$. Theorem 1 and 2 show that the configuration of $\mathcal U_k$ has an important effect on model capacity and generalization ability. On one hand, we need to make key users in $\mathcal U_k$ `representative' of diverse user behavior patterns on item consumption in order to guarantee enough representation capacity. Also, we need to control the size of $\mathcal U_k$ to maintain generalization ability. 

\subsection{Model Optimization}\label{sec-model3}
The complete training process is comprised of: \emph{pretraining} and \emph{adaption}. In the first stage, we train a MF model in transductive setting via Eq.~(\ref{eqn-obj1}) and obtain embeddings $\mathbf P_k$, $\mathbf Q$ and network $f_\theta$. The adaption involves optimization for the inductive relation model $h_w$ and finetuning the prediction network $f_\theta$ via Eq.~(\ref{eqn-obj3}). In particular, in terms of inductive learning for extrapolation (i.e., $\mathcal U_k = \mathcal U_q$), we further consider a self-supervised contrastive loss that pursuits similarity between the inductively computed user embeddings and the ones given by the MF model. Concretely, the loss is defined as
\begin{equation}
    \min_{w, \theta}\mathcal D_{\mathcal S_q}(\hat R_q, R_q) + \lambda \mathcal L_C(\mathbf P_k, \mathbf {\tilde P}_q),
\end{equation}
where $\lambda$ is a trading-off hyper-parameter and
\begin{equation}
    \mathcal L_C(\mathbf P_k, \mathbf {\tilde P}_q) = \frac{1}{M_q} \sum_{u}\log \left (\frac{\exp(\mathbf p_u^\top \mathbf {\tilde p}_{u})}{\sum_{u'\in \mathcal U_q}\exp(\mathbf p_u^\top \mathbf {\tilde p}_{u'})}\right ).
\end{equation}
The summation in the denominator can be approximated by in-batch negative samples with mini-batch training.

\begin{table*}[t!] 
\centering
\small
\caption{Statistics of five datasets used in our experiments. Amazon-Books and Amazon-Beauty datasets contain implicit user feedbacks while Douban, ML-100K, ML-1M have explicit feedbacks (ratings range within $[1,2,3,4,5]$).}
\label{tbl-dataset} 
\begin{tabular}{cccccccc}    
\toprule    
Dataset & \# Users & \#Items & \# Ratings & Density  & \# Key/Query Users & \# Training/Test Instances \\    
\midrule   
Douban & 3,000 & 3,000 & 0.13M & 0.0152  & 2,131/869 & 80,000/20,000 \\   
Movielens-100K & 943 & 1,682 & 0.10M & 0.0630 & 671/272 & 80,000/20,000 \\   
Movielens-1M & 6,040 & 3,706 & 1.0M & 0.0447  & 5,114/926 & 900,199/100,021 \\    
Amazon-Books & 52,643 & 91,599 & 2.1M & 0.0012 & 49,058/3,585 & 2,405,036/526,430 \\
Amazon-Beauty & 2,944 & 57,289 & 0.08M & 0.0004 & 780/2,164 & 53,464/29,440 \\
\bottomrule   
\end{tabular} 
\vspace{-10pt}
\end{table*}

\section{Comparison with Existing Works}\label{sec-comp}
We discuss related works and highlight our differences. 

\paragraph{Feature-driven Recommendation.} The collaborative filtering (CF) models do not assume any side information other than the rating matrix, but they cannot be trained in inductive ways due to the learnable user-specific embedding $\mathbf p_u$. To address the issue, previous works leverage side information, e.g. user profile features, to achieve inductive learning \cite{IMC, Feature1, Feature2, PinSage, NIMC}. Define user features (like age, occupation) as $\mathbf a_u$ and item features (like movie genre, director) as $\mathbf b_i$. The feature-driven model targets a prediction model $\hat r_{ui}=g(\mathbf a_u,\mathbf b_i)$. Since the space of $\mathbf a_u$ is shared among users, a model trained on one group of users can adapt to other users without retraining. However, feature-driven models often provide limited performance since the shared feature space is not expressive enough compared to user-specific embedding factors (see Appendix~\ref{appendix-express} for more discussions). Another issue is that high-quality features are hard to collect in practice. A key advantage of IDCF is the capability for inductive representation learning without using features.

\paragraph{Inductive Matrix Completion.} There are few existing works that attempt to handle inductive matrix completion using only user-item rating matrix. \cite{FEAE} (F-EAE) puts forward an exchangeable matrix layer that takes a whole rating matrix as input and inductively outputs prediction for missing ratings. However, the scalability of F-EAE is limited since it requires the whole rating matrix as input for training and inference for users, while IDCF enables mini-batch training and efficient inference. Besides, \cite{IGMC} (IGMC) proposes to use local subgraphs of user-item pairs in a bipartite graph of rating information as input features and further adopt graph neural networks to encode subgraph structures for rating prediction. The model achieves inductive learning via replacing users' one-hot indices by shared input features (i.e., index-free local subgraph structures). However, the expressiveness of IGMC is limited since the local subgraph structures can be indistinguishable for users with distinct behaviors (see Appendix~\ref{appendix-express} for more discussions), and the issue would become worse for implicit feedback data. By contrast, IDCF has equivalent expressiveness as original CF models. Another drawback of F-EAE and IGMC is that their models cannot output user representations. Differently, IDCF maintains the ability to give user-specific representations, which reflect users' preferences and can be used for downstream tasks (like user behavior modeling \cite{kalman}, social relation modeling \cite{featevo-wu}, user-controllable recommendation \cite{dis1}, target advertisement and influence maximization \cite{infmax1}, etc.).

\paragraph{Item-based CF Models.} Item-side observations contain rich information for use-side modeling and prediction \cite{dualseq-wu}. Previous works use item embeddings as user representation. \cite{item-based,FISM} adopt a combination of items rated by users to compute user embeddings and frees the model from learning parametric user-specific embeddings. Furthermore, there are quite a few auto-encoder architectures for recommendation problem, leveraging user's rating vector (ratings on all the items) as input, estimating user embedding (as latent variables), and decoding missing values in the rating vector \cite{AERec, MultVAE}. With item embeddings and user's rating history, these methods enable inductive learning for user representation and can adapt to new users on-the-fly. On methodological level, IDCF has the following differences. First, IDCF assumes learnable embeddings for both users and items, which maintains equivalent representation capacity as general MF/CF models (as proved in Theorem 1). Item-based models only consider learnable embeddings for items, and may suffer from limited representation capacity. Second, IDCF learns message-passing graphs among users by a relation model to obtain better user representations, instead of directly aggregating a given observed set of items' embeddings as item-based models.

\section{Experiment}\label{sec-exp}

In this section, we apply the proposed model IDCF to several real-world recommendation datasets to verify and dissect its effectiveness. Before going into the experiment results, we first introduce experiment setup including dataset information, evaluation protocol and implementation details.

\subsection{Experiment Setups}\label{sec-setup}

\paragraph{Datasets.} We consider five common recommendation benchmarks: \emph{Douban}, \emph{Movielens-100K (ML-100K)}, \emph{Movielens-1M (ML-1M)}, \emph{Amazon-Books} and \emph{Amazon-Beauty}. Douban, ML-100K and ML-1M have explicit user's ratings on movies. Amazon-Books and Amazon-Beauty contain implicit user feedbacks (the records of user's interactions with items). 
For Douban and ML-100K, we use the training/testing splits provided by \cite{sRGCNN}. For ML-1M\footnote{https://grouplens.org/datasets/movielens/}, we follow previous works \cite{GCMC, FEAE, IGMC} and use 9:1 training/testing spliting. For two Amazon datasets, we use the last ten interactions of each user for test and the remaining for training. We leave out $5\%$ training data as validation set for early stopping in training. Note that the raw datasets for Amazon-Books and Amazon-Beauty \footnote{http://jmcauley.ucsd.edu/data/amazon/} are very large and sparse ones and we filter out infrequent items and users with less than five ratings. The statistics of datasets used in our experiments are summarized in Table~\ref{tbl-dataset}. 

\begin{table*}[t!]
	\centering
	\caption{Test RMSE and NDCG for all the users (All) and few-shot users (FS) in Douban, ML-100K and ML-1M. We highlight the best scores among all the (resp. inductive) models with bold (resp. underline). \emph{Inductive} indicates whether the method can achieve inductive learning. \emph{Feature} indicates whether the method relies on user features. }
	\label{tbl-comp1}
	\footnotesize
	\begin{threeparttable}
\setlength{\tabcolsep}{1mm}{ 
		\begin{tabular}{ccccccccccccccc}
			\toprule[1pt]
			 \multirow{3}{*}{Method} & \multirow{3}{*}{Inductive} & \multirow{3}{*}{Feature} & \multicolumn{4}{c}{Douban}& \multicolumn{4}{c}{ML-100K}& \multicolumn{4}{c}{ML-1M} \\
            \cline{4-7} \cline{8-11} \cline{12-15}
            ~ & ~ & ~ & \multicolumn{2}{c}{RMSE}& \multicolumn{2}{c}{NDCG} & \multicolumn{2}{c}{RMSE}& \multicolumn{2}{c}{NDCG} & \multicolumn{2}{c}{RMSE}& \multicolumn{2}{c}{NDCG} \\
             \cline{4-5} \cline{6-7}   \cline{8-9}   \cline{10-11}  \cline{12-13}  \cline{14-15} 
		    ~& ~ & ~ &  All  & FS & All & FS  & All  & FS & All & FS & All  & FS & All & FS    \\ 
            \midrule[0.5pt]
			PMF & No & No & 0.737 & 0.718 & 0.939 & 0.954 & 0.932 & 1.003 & 0.858 & 0.843   & 0.851 & 0.946  & 0.919 & \textbf{0.940} \\
            NNMF & No & No & 0.729 & \textbf{0.705}& 0.939 & 0.952 & 0.925 & 0.987 & 0.895 & 0.878 & 0.848 & \textbf{0.940} &  0.920 & 0.937 \\
			GCMC  & No & No & 0.731 & 0.706 & 0.938 & \textbf{0.956} & 0.911 & 0.989 & 0.900 & \textbf{0.886} & \textbf{0.837}& 0.947 & 0.923 & 0.939  \\
            \midrule[0.5pt]
            NIMC & Yes & Yes & 0.732 &0.745 & 0.928 & 0.931 &1.015 &1.065 & 0.832  &  0.824
 & 0.873 & 0.995 & 0.889 & 0.904 \\  
            BOMIC & Yes & Yes  & 0.735 & 0.747 & 0.923  & 0.925 & 0.931 & 1.001 & 0.828 & 0.815 &0.847 & 0.953 & 0.905 & 0.924\\
            F-EAE  & Yes & No & 0.738 & - & - & -  & 0.920 & - & - & - &0.860 & - & - & - \\
			IGMC  & Yes & No & \underline{\textbf{0.721}} & 0.728 &  - & -  & \underline{\textbf{0.905}} & 0.997  & - & - & 0.857 & 0.956 & - & -   \\
			\textbf{IDCF-NN (ours)}  & Yes & No & 0.738 & \underline{0.712} & 0.939 & \underline{\textbf{0.956}} & 0.931 & 0.996 & 0.896 & 0.880 & 0.844 & 0.952 & 0.922 & \underline{\textbf{0.940}} \\
			\textbf{IDCF-GC (ours)} & Yes & No & 0.733 & \underline{0.712} & \underline{\textbf{0.940}} & \underline{\textbf{0.956}} & \underline{\textbf{0.905}} & \underline{\textbf{0.981}} &  \underline{\textbf{0.901}} & \underline{0.884} & \underline{0.839} & \underline{0.944} & \underline{\textbf{0.924}} & \underline{\textbf{0.940}} \\
			\hline
		\end{tabular}}
	\end{threeparttable}
\vspace{-10pt}
\end{table*}

\paragraph{Evaluation.} For datasets with explicit feedbacks, the goal is to predict user's ratings on items, i.e. estimate the missing values in user-item rating matrix. The task can be seen as a multi-class classification or regression problem. We use RMSE and NDCG to evaluate general reconstruction error and personalized ranking performance. RMSE counts the overall l2 distance from predicted ratings to the ground-truth, while NDCG is an averaged score that measures the consistency between the ranking of predicted ratings and that of the ground-truth for each user. For datasets with implicit feedbacks, the goal is to predict whether a user interacts with an item. The task is essentially a one-class classification problem. Since the dataset is very sparse and only has positive instances, we uniformly sample five items as negative samples for each clicked item and adopt AUC and NDCG to measure the global and personalized ranking accuracy, respectively. AUC is short of Area Under of the ROC Curve which measures the global consistency between the ranking of all the predicted user-item interactions and the ground-truth (which ranks all the 1's before 0'). More details for evaluation metrics are provided in Appendix~\ref{appendix-metric} 

\paragraph{Implementations.} We consider two specifications for $f_\theta$ in our model: IDCF-NN, which adopts multi-layer perceptron for $f$, and IDCF-GC, which uses graph convolution network for $f$.

\textit{Feedforward Neural Network as Matrix Factorization Model (IDCF-NN).}
We follow the architecture in NNMF \cite{NNMF} and use neural network for $f_\theta$. Here we combine a three-layer neural network and a shallow dot-product operation. Concretely,
\begin{equation}\label{eqn-C1-model}
    f_\theta(\mathbf p_u, \mathbf q_i) = \frac{(\mathbf p_u^\top \mathbf q_i + nn([\mathbf p_u \| \mathbf q_i \| \mathbf p_u \odot \mathbf q_i])) }{2} + b_u + b_i,
\end{equation}
where $nn$ is a three-layer neural network using $tanh$ activation, $\odot$ denotes element-wise product and $b_u$, $b_i$ are bias terms for user $u$ and item $i$, respectively. 

\textit{Graph Convolution Network as Matrix Factorization Model (IDCF-GC).}
We follow the architecture in GCMC \cite{GCMC} and adopt graph convolution network for $f_\theta()$. Besides user-specific embedding for user $u$ and item-specific embedding for item $i$, we consider embeddings for user $u$'s rated items and users who rated on item $i$, i.e., the one-hop neighbors of $u$ and $i$ in user-item bipartite graph. Denote $\mathcal N_{u, m}=\{i| r_{ui}= m\}$ as user $u$'s rated items with rating value $m$ and $\mathcal N_{i,m}=\{u| r_{ui}=m\}$ as users who rated on item $i$ with rating value $m$ for $m\geq 1$. Consider graph convolution to aggregate information from neighbors,
\begin{equation}
\begin{aligned}
    \mathbf m_{u, m} =& ReLU(\frac{1}{|\mathcal N_{u,m}|} \sum_{i\in \mathcal N_{u,m}} \mathbf W_{q, m} \mathbf q_i), \\
    \mathbf n_{i, m} =& ReLU(\frac{1}{|\mathcal N_{i, m}|} \sum_{u\in \mathcal N_{i, m}} \mathbf W_{p, m} \mathbf p_u),
    \end{aligned}
\end{equation}
and combination function: $\mathbf m_u = \mbox{FC}(\{\mathbf m_{u,m}\}_{m})$, $\mathbf n_i = \mbox{FC'}(\{\mathbf n_{i,m}\}_{m})$ where $\mbox{FC}$ denotes a fully-connected layer. Then we define the output function
\begin{equation}\label{eqn-C2-model}
\begin{split}
     &f(\mathbf p_u, \mathbf q_i, \{\mathbf p_u\}_{u\in \mathcal N_i}, \{\mathbf q_i\}_{i\in \mathcal N_u}) \\
     =& nn'([\mathbf p_u \odot \mathbf q_i \| \mathbf p_u \odot \mathbf m_u \| \mathbf n_i \odot \mathbf q_i \| \mathbf n_i \odot \mathbf m_u]) + b_u + b_i,   
\end{split}
\end{equation}
where $nn'$ is a three-layer neural network using $ReLU$ activation function. In Appendix~\ref{appendix-paras}, we provide more details for hyper-parameter setting. Moreover, we specify $l(\hat r_{ui}, r_{ui})$ as MSE loss (resp. cross-entropy) for explicit (resp. implicit) feedback data.




\subsection{Comparative Results}\label{sec-exp-comp}

\subsubsection{Interpolation for Few-Shot Users.}\label{sec-exp-comp1}

\textbf{Setup.} We study the performance on few-shot query users with limited training ratings. We split users in each dataset into two sets: users with more than $\delta$ training ratings, denoted as $ \mathcal{U}_1$, and those with less than $\delta$ training ratings $\mathcal U_2$. We basically set $\delta=30$ for Douban, ML-100K and ML-1M datasets, and $\delta=20$ for Amazon datasets. For IDCF, we adopt the training algorithm of \emph{inductive learning for interpolation} with $ \mathcal{U}_1$ as key users and $ \mathcal{U}_2$ as query users. We use the training ratings for $ \mathcal{U}_1$ and $ \mathcal{U}_2$ as input data for our pretraining and adaption, respectively. We compare with several competitors, including 1) powerful transductive methods PMF \cite{PMF}, NNMF \cite{NNMF} and GCMC \cite{GCMC}, 2) inductive feature-driven methods NIMC \cite{NIMC}, BOMIC \cite{BOMIC}, and 3) inductive matrix completion model IGMC \cite{IGMC}. We train each competitor with training ratings of $\mathcal{U}_1$ and $ \mathcal{U}_2$.

Table~\ref{tbl-comp1} reports RMSE and NDCG for test ratings of all the users and few-shot users in $\mathcal U_2$ on Douban, ML-100k and ML-1M. As we can see, IDCF-NN (resp. IDCF-GC) gives very close RMSE and NDCG on all the users and query users to NNMF (resp. GCMC), which suggests that our inductive model can achieve similar reconstruction error and ranking accuracy as corresponding transductive counterpart. The results validate our theoretical analysis in Section~\ref{sec-model} that IDCF possesses the same representation capacity as matrix factorization model. Even though IDCF enables inductive representation learning via a parameter-sharing relation model, it does not sacrifice any representation power. Compared with inductive methods, IDCF-GC achieves the best RMSE and NDCG for query users in most cases. The results demonstrate the superiority of IDCF against other feature-driven and inductive matrix completion models. 

Table~\ref{tbl-comp2} shows AUC and NDCG for test interactions of few-shot users in $\mathcal U_2$ on Amazon-Books and Amazon-Beauty. Since Amazon datasets have no user feature, the feature-driven competitors would not work. Also, IGMC would fail to work with implicit feedbacks. We compare with two other graph-based recommendation models, NGCF \cite{ngcf} and PinSAGE \cite{PinSage}. As shown in Table~\ref{tbl-comp2}, IDCF-NN and IDCF-GC significantly outperform transductive models in implicit feedback setting with $2.3\%$/$3.0\%$ (resp. $1.0\%$/$1.0\%$) improvement of AUC/NDCG on Amazon-Books (resps. Amazon-Beauty). 
The two Amazon datasets are both very sparse with rating density approximately $0.001$. One implication here is that our inductive model can provide better performance than transductive models for users with few training ratings.

\begin{table}[t!]
	\centering
	\caption{Test AUC and NDCG for few-shot users (FS) and new users (New) in Amazon-Books and Amazon-Beauty.}
	\label{tbl-comp2}
	\footnotesize        	
	\begin{threeparttable}
		\resizebox{0.48\textwidth}{!}{
\setlength{\tabcolsep}{1mm}{ 
		\begin{tabular}{ccccccccc}
			\toprule[1pt]
        \multirow{2}{*}{Method}  & \multicolumn{4}{c}{Amazon-Books} &  \multicolumn{4}{c}{Amazon-Beauty} \\
         \cline{2-9} 
        ~ & \multicolumn{2}{c}{AUC} & \multicolumn{2}{c}{NDCG}     & \multicolumn{2}{c}{AUC}     & \multicolumn{2}{c}{NDCG}  \\
        \cline{2-9} 
        ~ & Query & New & Query & New  & Query & New & Query & New      \\
        \midrule[0.4pt]
        PMF & 0.917 & - & 0.888 & - & 0.779 & - & 0.769 & -      \\
        NNMF & 0.919     & - & 0.891 & - & 0.790 & - & 0.763 & -      \\
        NGCF &  0.916    &  - & 0.896 & - & 0.793 & - & 0.775 & -      \\
        PinSAGE &  0.923    &  - & 0.901 & - & 0.790 & - & 0.775 & -      \\
        FISM & -     &  0.752 & - & 0.792 & - & 0.613 & - & 0.678      \\
        MultVAE &  -    &  0.738 & - & 0.701 & - & 0.644 & - & 0.679      \\
        \textbf{IDCF-NN} & \textbf{0.944}     &  0.939 & \textbf{0.928} & 0.920 & 0.792 & 0.750
 & \textbf{0.783} & 0.774      \\
        \textbf{IDCF-GC} &  0.938  &  \textbf{0.946} & 0.921 & \textbf{0.930 } & \textbf{0.801} & \textbf{0.791} & 0.772 & \textbf{0.791}      \\
        \bottomrule[1pt]
		\end{tabular}}}
	\end{threeparttable}
\vspace{-15pt}
\end{table}

\begin{table}[t!]
	\centering
	\caption{Test RMSE and NDCG for new users on Douban, ML-100K and ML-1M.}
	\label{tbl-comp3}
	\footnotesize
	\begin{threeparttable}
\setlength{\tabcolsep}{1mm}{ 
		\begin{tabular}{ccccccc}
			\toprule[1pt]
        \multirow{2}{*}{Method}   & \multicolumn{2}{c}{Douban} & \multicolumn{2}{c}{ML-100K} & \multicolumn{2}{c}{ML-1M} \\
        \cline{2-7}
        ~ & RMSE & NDCG  & RMSE & NDCG   & RMSE & NDCG  \\
        \midrule[0.4pt]
        NIMC &  0.766  & 0.921 & 1.089 & 0.864  & 1.059 & 0.883   \\
        BOMIC & 0.764   & 0.920 & 1.088 & 0.859 & 1.057 &  0.879 \\
        FISM & 1.910  & 0.824  & 1.891  & 0.760 & 2.283 & 0.771 \\
        MultVAE & 2.783  & 0.823  & 2.865  & 0.758  & 2.981 & 0.792   \\
        IGMC &  0.743 & -  & 1.051  & - &  0.997  &-  \\
        \textbf{IDCF-NN} & 0.749  & \textbf{0.955} & 1.078 & 0.877 & 0.994 & 0.941    \\
        \textbf{IDCF-GC} &  \textbf{0.723}  & \textbf{0.955} &  \textbf{1.011} & \textbf{0.881} & \textbf{0.957} &  \textbf{0.942}   \\
        \bottomrule[1pt]
		\end{tabular}}
	\end{threeparttable}
\vspace{-15pt}
\end{table}

\begin{figure*}[t!]
\begin{minipage}{0.99\linewidth}
\centering
\subfigure[]{
\begin{minipage}[t]{0.19\linewidth}
\centering
\label{fig-ratio}
\includegraphics[width=0.95\textwidth,angle=0]{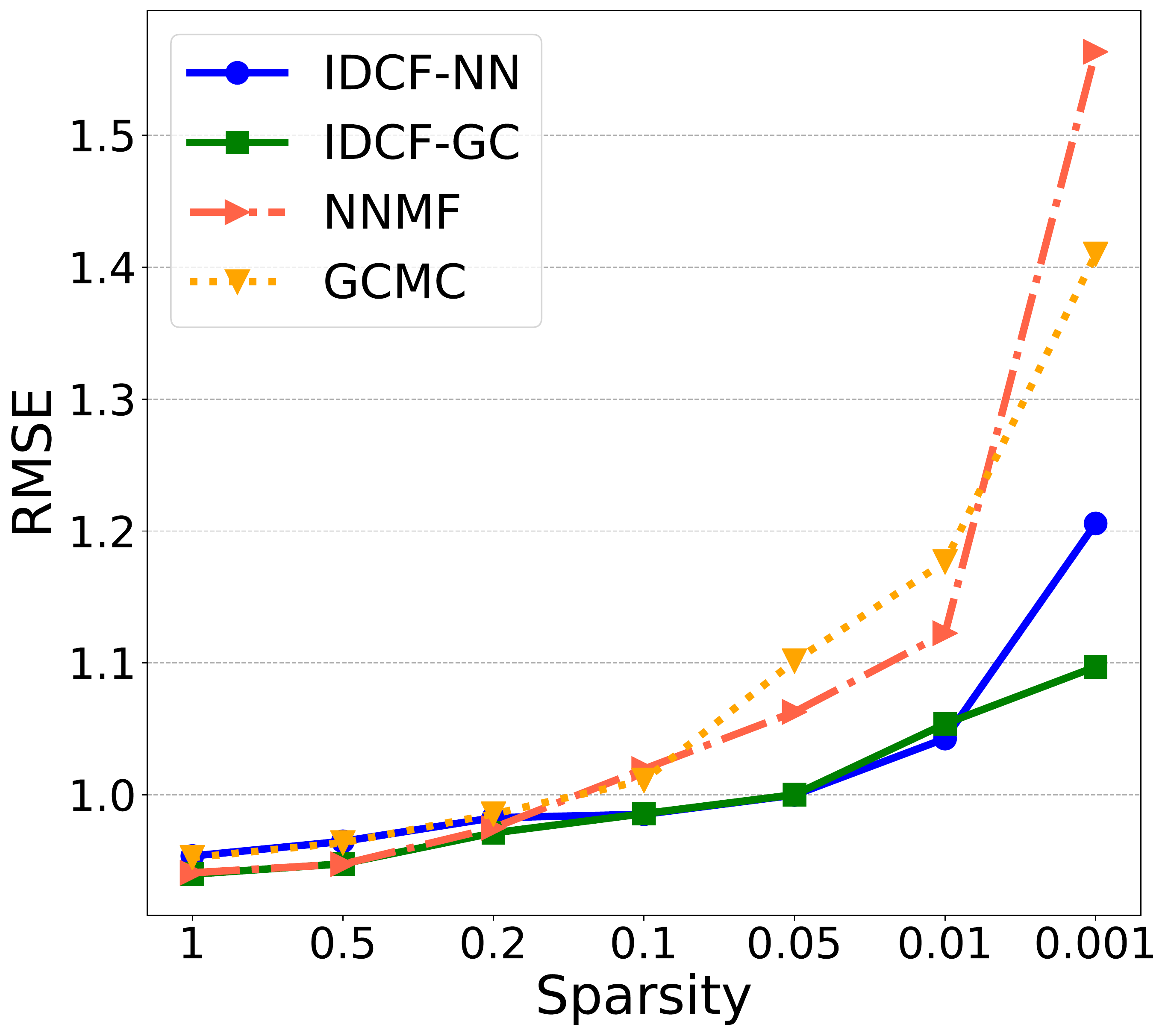}
\end{minipage}%
}%
\subfigure[]{
\begin{minipage}[t]{0.19\linewidth}
\centering
\label{fig-sparse}
\includegraphics[width=0.95\textwidth,angle=0]{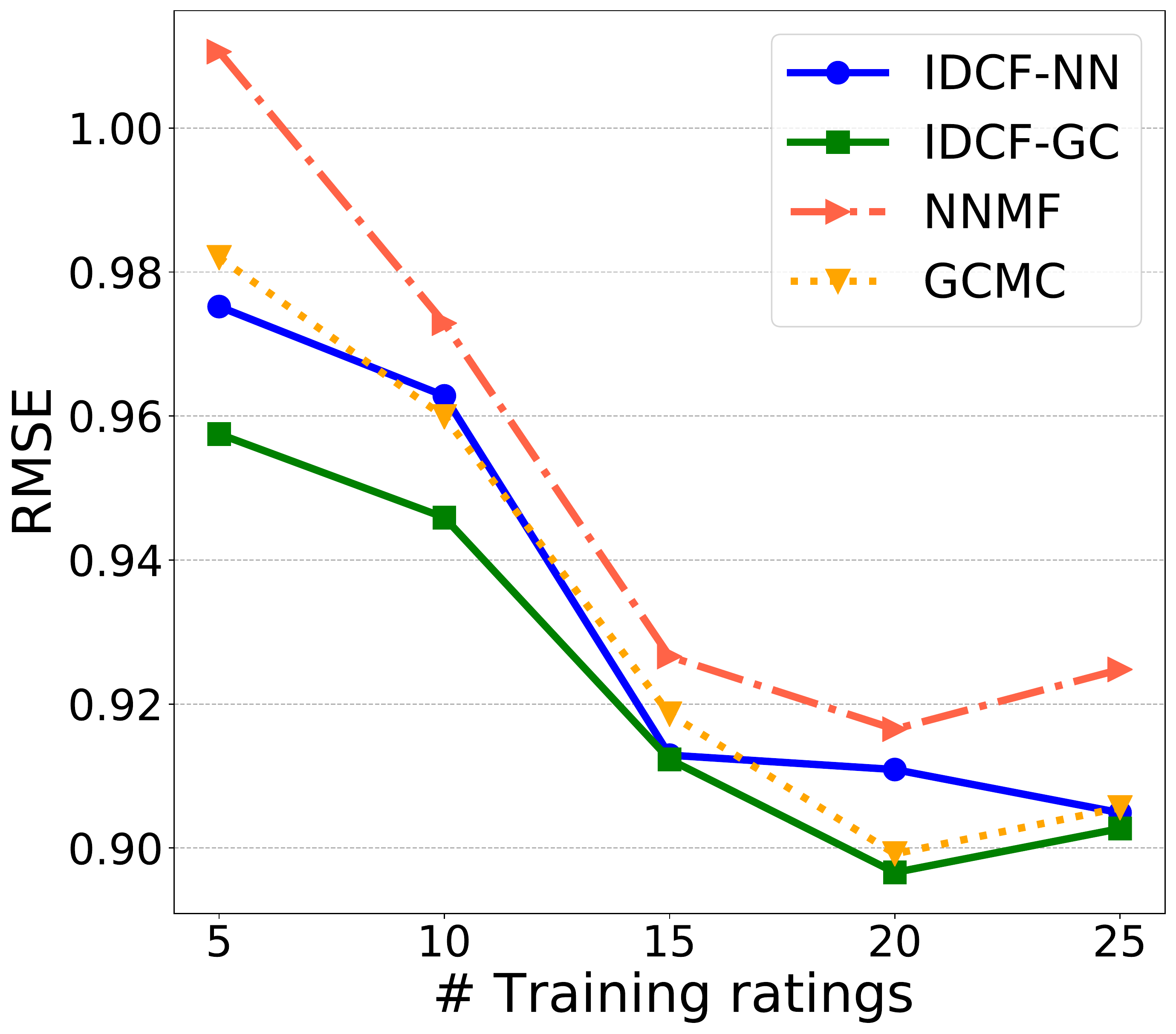}
\end{minipage}%
}%
\subfigure[]{
\begin{minipage}[t]{0.2\linewidth}
\centering
\label{fig-attn-a}
\includegraphics[width=0.91\textwidth,angle=0]{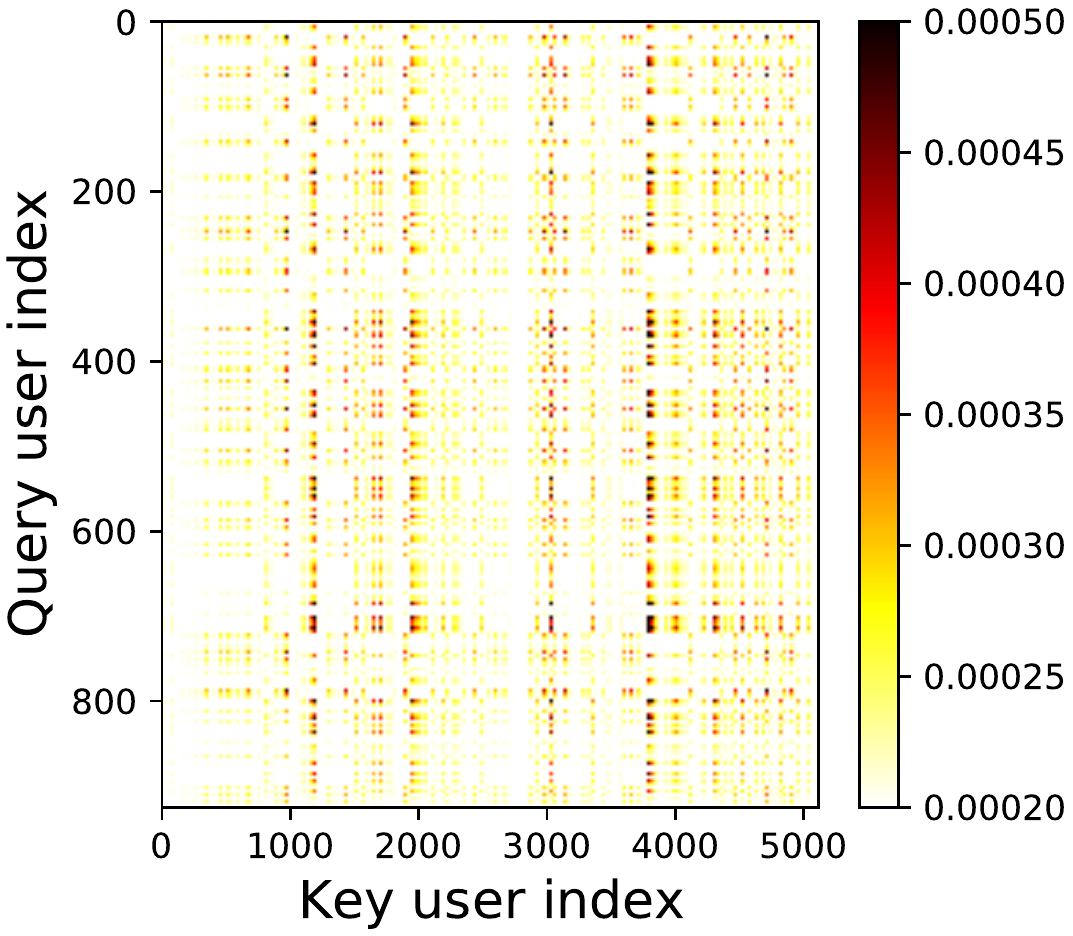}
\end{minipage}%
}%
\subfigure[]{
\begin{minipage}[t]{0.2\linewidth}
\centering
\label{fig-attn-b}
\includegraphics[width=0.89\textwidth,angle=0]{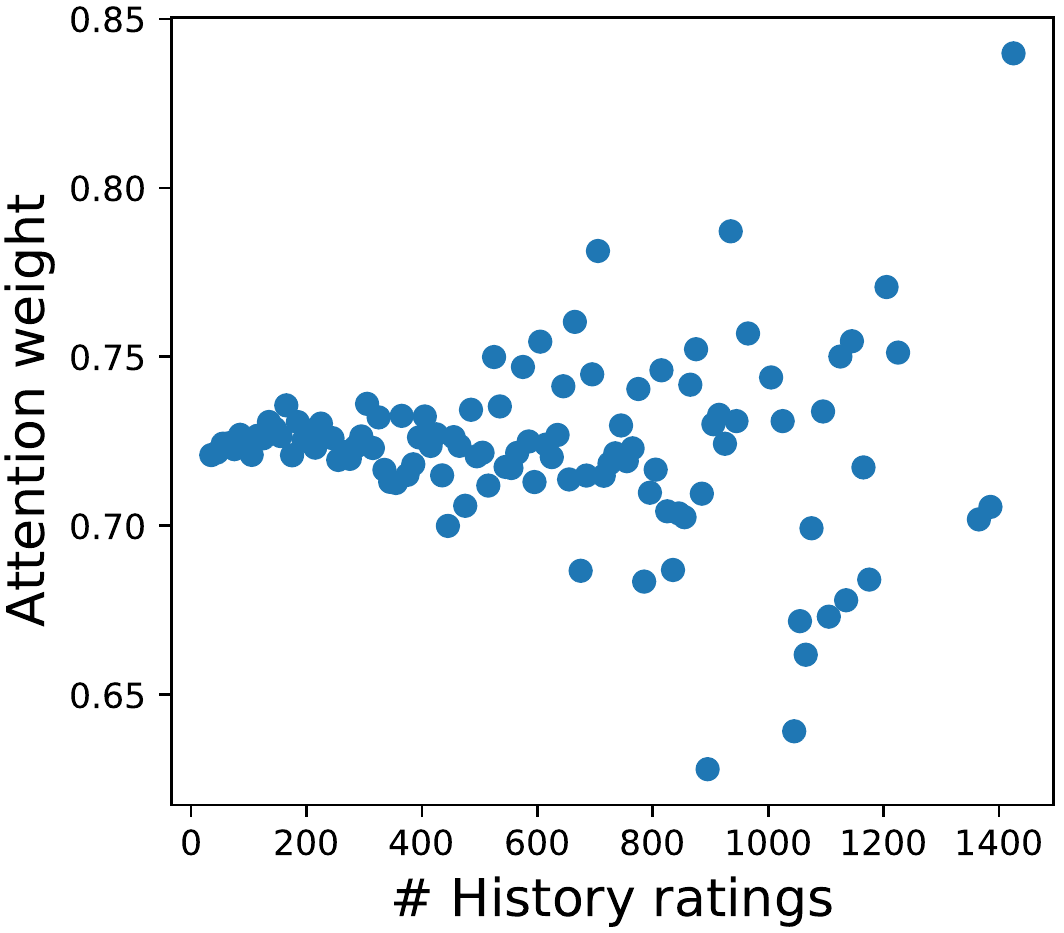}
\vspace{2pt}
\end{minipage}%
}%
\subfigure[]{
\begin{minipage}[t]{0.19\linewidth}
\centering
\label{fig-sca}
\includegraphics[width=0.96\textwidth,angle=0]{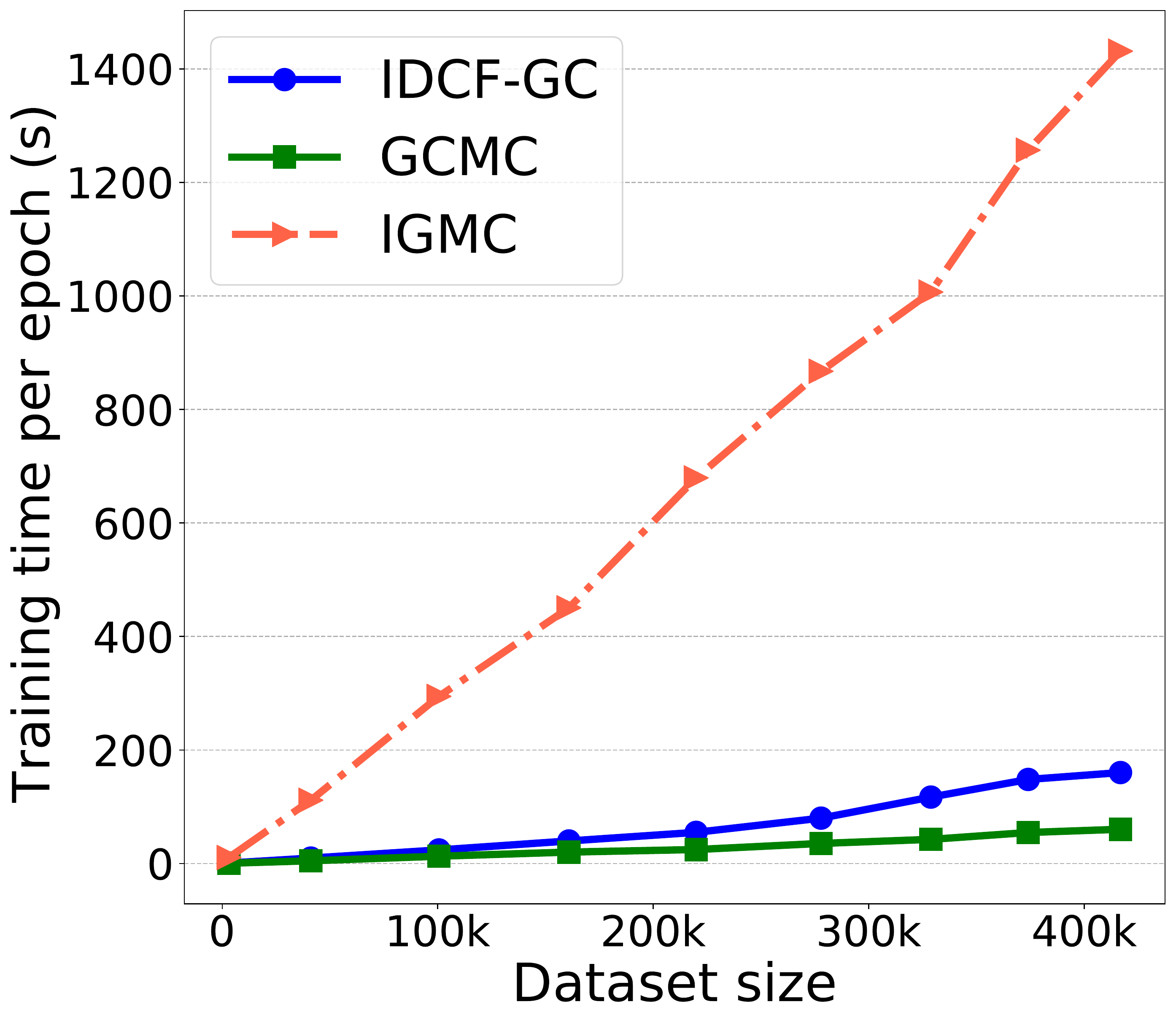}
\end{minipage}
}
\end{minipage}
\vspace{-10pt}
\caption{Evaluation. (a) Overall RMSE w.r.t \# sparsity ratio. (b) User-specific RMSE w.r.t \# user's training ratings. (c) Attention weights of query users (y-axis) on key users (x-axis). (d) Key uses' accumulated attention weights w.r.t. \# historical ratings. (e) Scalability test.}
\vspace{-10pt}
\end{figure*}
\subsubsection{Extrapolation for New Users.}

\textbf{Setup.} We then investigate model's generalization performance on new users that are unseen in training. We assume the model is only exposed to the training ratings of $\mathcal U_1$ and test its performance on test ratings of $\mathcal U_2$. Concretely, for IDCF, we leverage the training algorithm of \emph{inductive learning for extrapolation} with $\mathcal U_1$ as both key and query users. The two-stage training uses training ratings of $\mathcal U_1$. We compare with inductive models NIMC, BOMIC, IGMC and item-based CF models FISM \cite{FISM}, MultVAE \cite{MultVAE}.

Table~\ref{tbl-comp3} reports results on test ratings for new users in $\mathcal U_2$ on Douban, ML-100K and ML-1M. Notably,  IDCF-GC outperforms the best competitors by a large margin, with RMSE (resp. NDCG) improvement of $2.6\%$ (resp. $3.6\%$) on Douban, $3.8\%$ (resp. $2.0\%$) on ML-100K and $4.0\%$ (resp. $6.7\%$) on ML-1M. Also, Table~\ref{tbl-comp2} reports test AUC and NDCG for new users on two Amazon datasets. The results show that both IDCF-NN and IDCF-GC outperform other competitors with $25.7\%$ (resp. $17.4\%$) and $22.8\%$ (resp. $16.4\%$) improvements of AUC (resp. NDCG) on Amazon-Books and Amazon-Beauty, respectively. Such results demonstrate superior power of IDCF for addressing new users in open-world recommendation.



\subsection{Further Discussion}\label{sec-exp-dis}


\paragraph{Sparse Data and Few-shot Users.} A successful recommender system is supposed to handle data sparsity and few-shot users with few historical ratings. Here we construct  sparse datasets by using $50\%, 20\%, 10\%, 5\%, 1\%$ and $0.1\%$ training ratings in Movielens-1M, and then compare the test RMSEs of query users in Fig.~\ref{fig-ratio}. Also, Fig.~\ref{fig-sparse} compares the test RMSEs for users with different numbers of historical ratings under $50\%$ sparsity. As shown in Fig.~\ref{fig-ratio}, as the dataset becomes sparser, the RMSEs of all the models suffer from a drop, but the drop rate of our inductive models IDCF-NN and IDCF-GC is much smaller compared with transductive models NNMF and GCMC. In Fig.~\ref{fig-sparse}, we find that users with more historical ratings usually have better RMSE scores compared with few-shot users. By contrast, our inductive models IDCF-NN and IDCF-GC exhibit a more smooth decrease and even outperform other transductive methods NNMF and GCMC for users with very few ratings. In the extreme cases with less than five historical ratings, IDCF-GC achieves $2.5\%$ improvement on RMSE compared with the best transductive method GCMC.


\textbf{Attention Weight Distribution} In Fig.~\ref{fig-attn-a} we visualize attention weights of IDCF-NN from query users to key users in ML-1M. There is an interesting phenomenon that some of key users appear to be very `important' and most query users have high attention scores on them. It indicates that the embeddings of these key users are informative and can provide powerful expressiveness for query users' preferences. In Fig.~\ref{fig-attn-b} we further plots key users' accumulated attention weights (sum of the attention scores over all the query users) w.r.t. \# historical ratings. We can see that key users with more historical ratings are more likely to have large attention weights on query users, though they are also more likely to have low attention scores. This observation is consistent with intuition that the users with more historical ratings are easier for the model to identify their interests. Also, the results gives an important hint for selecting useful key users: informative key users are more likely to exsit in users with more historical ratings. In Appendix~\ref{appendix-moreres}, we compare different split ways for key and query users and will provide more discussions on this point.




\textbf{Scalability Test}
We further investigate the scalability of IDCF-GC compared with two GNN-based counterparts IGMC and GCMC. We statistic the training time per epoch on ML-1M using a GTX 1080Ti with 11G memory. Here we truncate the dataset and use different numbers of ratings for training. 
The results are shown in Fig.~\ref{fig-sca}.
As we can see, when dataset size becomes large, the training times per epoch of three models all exhibit linear increase. IDCF spends approximately one more time than GCMC, while IGMC is approximately ten times slower than IDCF. 
Nevertheless, while IDCF costs one more training time than GCMC, the latter cannot tackle new unseen users without retraining a model in test stage. 

 
\textbf{More Insights on IDCF's Effectiveness.} The relation model aims at learning graph structures among users that explore more useful proximity information and maximize the benefits of message passing for inductive representation learning. Existing graph-based (recommendation) models rely on a given observed graph, on top of which a GNN model would play as a strong inductive bias. Such inductive bias may be beneficial for some cases (if the graph is fully observed and possesses homophily property) and harmful for other cases (if the graph has noisy/missing links or have heterophily structures). Differently, IDCF enables mutual reinforcement between structure learning and message passing with the guidance of supervised signals from downstream tasks in a fully data-driven manner.

\section{Conclusions and Outlook}
We have proposed an inductive collaborative filtering framework that learns hidden relational graphs among users to allow effective message passing in the latent space. It accomplishes inductive computation for user-specific representations without compromising on representation capacity and scalablity. Our model achieves state-of-the-art performance on inductive collaborative filtering for recommendation with few-shot and zero-shot users that are commonly encountered in open-world recommendation.

The core idea of IDCF opens a new way for next generation of representation learning, i.e., one can consider a pretrained representation model for one set of existing entities and their representations (through some simple transformations) can be generalized to efficiently compute inductive representations for others, enabling the model to flexibly handle new coming entities in the wild. 


\section*{Acknowledgement}

We would like to thank the anonymous reviewers for their valuable feedbacks and suggestions that help to improve this work. This work was supported by the National Key R\&D Program of China [2020YFB1707903], the National Natural Science Foundation of China [61872238,61972250], Shanghai Municipal Science and Technology Major Project (2021SHZDZX0102), the Tencent Marketing Solution Rhino-Bird Focused Research Program [FR202001], and the CCF-Tencent Open Fund [RAGR20200105].


{
\bibliography{icml2021_conference}

\begin{thebibliography}{37}
\providecommand{\natexlab}[1]{#1}
\providecommand{\url}[1]{\texttt{#1}}
\expandafter\ifx\csname urlstyle\endcsname\relax
  \providecommand{\doi}[1]{doi: #1}\else
  \providecommand{\doi}{doi: \begingroup \urlstyle{rm}\Url}\fi

\bibitem[Cheng et~al.(2016)Cheng, Koc, Harmsen, Shaked, Chandra, Aradhye,
  Anderson, Corrado, Chai, Ispir, Anil, Haque, Hong, Jain, Liu, and
  Shah]{Feature2}
Cheng, H., Koc, L., Harmsen, J., Shaked, T., Chandra, T., Aradhye, H.,
  Anderson, G., Corrado, G., Chai, W., Ispir, M., Anil, R., Haque, Z., Hong,
  L., Jain, V., Liu, X., and Shah, H.
\newblock Wide {\&} deep learning for recommender systems.
\newblock In \emph{DLRS}, pp.\  7--10, 2016.

\bibitem[Covington et~al.(2016)Covington, Adams, and Sargin]{youtube}
Covington, P., Adams, J., and Sargin, E.
\newblock Deep neural networks for youtube recommendations.
\newblock In \emph{RecSys}, pp.\  191--198, 2016.

\bibitem[Cremonesi et~al.(2010)Cremonesi, Koren, and Turrin]{item-based}
Cremonesi, P., Koren, Y., and Turrin, R.
\newblock Performance of recommender algorithms on top-n recommendation tasks.
\newblock In \emph{RecSys}, pp.\  39--46. {ACM}, 2010.

\bibitem[Dziugaite \& Roy(2015)Dziugaite and Roy]{NNMF}
Dziugaite, G.~K. and Roy, D.~M.
\newblock Neural network matrix factorization.
\newblock \emph{CoRR}, abs/1511.06443, 2015.

\bibitem[Gilmer et~al.(2017)Gilmer, Schoenholz, Riley, Vinyals, and
  Dahl]{chemi-messpass}
Gilmer, J., Schoenholz, S.~S., Riley, P.~F., Vinyals, O., and Dahl, G.~E.
\newblock Neural message passing for quantum chemistry.
\newblock In \emph{ICML}, pp.\  1263--1272, 2017.

\bibitem[Hamilton et~al.(2017)Hamilton, Ying, and Leskovec]{graphsage}
Hamilton, W.~L., Ying, Z., and Leskovec, J.
\newblock Inductive representation learning on large graphs.
\newblock In \emph{NeurIPS}, pp.\  1024--1034, 2017.

\bibitem[Hartford et~al.(2018)Hartford, Graham, Leyton{-}Brown, and
  Ravanbakhsh]{FEAE}
Hartford, J.~S., Graham, D.~R., Leyton{-}Brown, K., and Ravanbakhsh, S.
\newblock Deep models of interactions across sets.
\newblock In \emph{ICML}, pp.\  1914--1923, 2018.

\bibitem[Hu et~al.(2008)Hu, Koren, and Volinsky]{Imp-early}
Hu, Y., Koren, Y., and Volinsky, C.
\newblock Collaborative filtering for implicit feedback datasets.
\newblock In \emph{ICDM}, pp.\  263--272, 2008.

\bibitem[Jain \& Dhillon(2013)Jain and Dhillon]{IMC}
Jain, P. and Dhillon, I.~S.
\newblock Provable inductive matrix completion.
\newblock \emph{CoRR}, abs/1306.0626, 2013.

\bibitem[Kabbur et~al.(2013)Kabbur, Ning, and Karypis]{FISM}
Kabbur, S., Ning, X., and Karypis, G.
\newblock {FISM:} factored item similarity models for top-n recommender
  systems.
\newblock In \emph{KDD}, pp.\  659--667. {ACM}, 2013.

\bibitem[Kang et~al.(2020)Kang, Cheng, Chen, Yi, Lin, Hong, and
  Chi]{quantized-emb}
Kang, W., Cheng, D.~Z., Chen, T., Yi, X., Lin, D., Hong, L., and Chi, E.~H.
\newblock Learning multi-granular quantized embeddings for large-vocab
  categorical features in recommender systems.
\newblock In \emph{WWW}, pp.\  562--566, 2020.

\bibitem[Khalil et~al.(2017)Khalil, Dai, Zhang, Dilkina, and Song]{infmax1}
Khalil, E.~B., Dai, H., Zhang, Y., Dilkina, B., and Song, L.
\newblock Learning combinatorial optimization algorithms over graphs.
\newblock In \emph{NeurIPS}, pp.\  6348--6358, 2017.

\bibitem[Koren et~al.(2009)Koren, Bell, and Volinsky]{MFrec}
Koren, Y., Bell, R., and Volinsky, C.
\newblock Matrix factorization techniques for recommender systems.
\newblock \emph{Computer}, 42\penalty0 (8):\penalty0 30--37, 2009.

\bibitem[Ledent et~al.(2020)Ledent, Alves, and Kloft]{BOMIC}
Ledent, A., Alves, R., and Kloft, M.
\newblock Orthogonal inductive matrix completion.
\newblock \emph{CoRR}, abs/2004.01653, 2020.

\bibitem[Lee et~al.(2019)Lee, Im, Jang, Cho, and Chung]{MeLU}
Lee, H., Im, J., Jang, S., Cho, H., and Chung, S.
\newblock Melu: Meta-learned user preference estimator for cold-start
  recommendation.
\newblock In \emph{KDD}, pp.\  1073--1082, 2019.

\bibitem[Liang et~al.(2018)Liang, Krishnan, Hoffman, and Jebara]{MultVAE}
Liang, D., Krishnan, R.~G., Hoffman, M.~D., and Jebara, T.
\newblock Variational autoencoders for collaborative filtering.
\newblock In Champin, P., Gandon, F.~L., Lalmas, M., and Ipeirotis, P.~G.
  (eds.), \emph{WWW}, pp.\  689--698. {ACM}, 2018.

\bibitem[Liu et~al.(2020)Liu, Lu, Zhao, Xu, Peng, Liu, Zhang, Li, Jin, Bao, and
  Yan]{kalman}
Liu, H., Lu, J., Zhao, X., Xu, S., Peng, H., Liu, Y., Zhang, Z., Li, J., Jin,
  J., Bao, Y., and Yan, W.
\newblock Kalman filtering attention for user behavior modeling in {CTR}
  prediction.
\newblock In \emph{NeurIPS}, 2020.

\bibitem[Ma et~al.(2019)Ma, Zhou, Cui, Yang, and Zhu]{dis1}
Ma, J., Zhou, C., Cui, P., Yang, H., and Zhu, W.
\newblock Learning disentangled representations for recommendation.
\newblock In \emph{NeurIPS}, pp.\  5712--5723, 2019.

\bibitem[Mikolov et~al.(2013)Mikolov, Chen, Corrado, and Dean]{word2vec}
Mikolov, T., Chen, K., Corrado, G., and Dean, J.
\newblock Efficient estimation of word representations in vector space.
\newblock In \emph{ICLR, Workshop}, 2013.

\bibitem[Mohri et~al.(2012)Mohri, Rostamizadeh, and Talwalkar]{foundation}
Mohri, M., Rostamizadeh, A., and Talwalkar, A.
\newblock \emph{Foundations of Machine Learning}.
\newblock Adaptive computation and machine learning. {MIT} Press, 2012.

\bibitem[Monti et~al.(2017)Monti, Bronstein, and Bresson]{sRGCNN}
Monti, F., Bronstein, M.~M., and Bresson, X.
\newblock Geometric matrix completion with recurrent multi-graph neural
  networks.
\newblock In Guyon, I., von Luxburg, U., Bengio, S., Wallach, H.~M., Fergus,
  R., Vishwanathan, S. V.~N., and Garnett, R. (eds.), \emph{NeurIPS}, pp.\
  3697--3707, 2017.

\bibitem[Qian et~al.(2019)Qian, Liang, and Li]{CSR-GNN}
Qian, T., Liang, Y., and Li, Q.
\newblock Solving cold start problem in recommendation with attribute graph
  neural networks.
\newblock \emph{CoRR}, abs/1912.12398, 2019.

\bibitem[Rendle et~al.(2009)Rendle, Freudenthaler, Gantner, and
  Schmidt{-}Thieme]{BPR}
Rendle, S., Freudenthaler, C., Gantner, Z., and Schmidt{-}Thieme, L.
\newblock {BPR:} bayesian personalized ranking from implicit feedback.
\newblock In \emph{{UAI}}, pp.\  452--461, 2009.

\bibitem[Salakhutdinov \& Mnih(2007)Salakhutdinov and Mnih]{PMF}
Salakhutdinov, R. and Mnih, A.
\newblock Probabilistic matrix factorization.
\newblock In Platt, J.~C., Koller, D., Singer, Y., and Roweis, S.~T. (eds.),
  \emph{NeurIPS}, pp.\  1257--1264, 2007.

\bibitem[Scarselli et~al.(2009)Scarselli, Gori, Tsoi, Hagenbuchner, and
  Monfardini]{GNN-early1}
Scarselli, F., Gori, M., Tsoi, A.~C., Hagenbuchner, M., and Monfardini, G.
\newblock The graph neural network model.
\newblock \emph{{IEEE} Trans. Neural Networks}, 20\penalty0 (1):\penalty0
  61--80, 2009.

\bibitem[Sedhain et~al.(2015)Sedhain, Menon, Sanner, and Xie]{AERec}
Sedhain, S., Menon, A.~K., Sanner, S., and Xie, L.
\newblock Autorec: Autoencoders meet collaborative filtering.
\newblock In \emph{WWW (Companion Volume)}, pp.\  111--112. {ACM}, 2015.

\bibitem[Srebro et~al.(2004)Srebro, Alon, and Jaakkola]{MFGenError}
Srebro, N., Alon, N., and Jaakkola, T.~S.
\newblock Generalization error bounds for collaborative prediction with
  low-rank matrices.
\newblock In \emph{NeurIPS}, pp.\  1321--1328, 2004.

\bibitem[van~den Berg et~al.(2017)van~den Berg, Kipf, and Welling]{GCMC}
van~den Berg, R., Kipf, T.~N., and Welling, M.
\newblock Graph convolutional matrix completion.
\newblock \emph{CoRR}, abs/1706.02263, 2017.

\bibitem[Wang et~al.(2019)Wang, He, Wang, Feng, and Chua]{ngcf}
Wang, X., He, X., Wang, M., Feng, F., and Chua, T.
\newblock Neural graph collaborative filtering.
\newblock In \emph{SIGIR}, pp.\  165--174, 2019.

\bibitem[Wu et~al.(2019{\natexlab{a}})Wu, Gao, Gao, Weng, and Chen]{dualseq-wu}
Wu, Q., Gao, Y., Gao, X., Weng, P., and Chen, G.
\newblock Dual sequential prediction models linking sequential recommendation
  and information dissemination.
\newblock In \emph{SIGKDD}, pp.\  447--457, 2019{\natexlab{a}}.

\bibitem[Wu et~al.(2019{\natexlab{b}})Wu, Jiang, Gao, Yang, and
  Chen]{featevo-wu}
Wu, Q., Jiang, L., Gao, X., Yang, X., and Chen, G.
\newblock Feature evolution based multi-task learning for collaborative
  filtering with social trust.
\newblock In \emph{IJCAI}, pp.\  3877--3883, 2019{\natexlab{b}}.

\bibitem[Wu et~al.(2019{\natexlab{c}})Wu, Zhang, Gao, He, Weng, Gao, and
  Chen]{dancer-wu}
Wu, Q., Zhang, H., Gao, X., He, P., Weng, P., Gao, H., and Chen, G.
\newblock Dual graph attention networks for deep latent representation of
  multifaceted social effects in recommender systems.
\newblock In \emph{The Web Conference}, pp.\  2091--2102, 2019{\natexlab{c}}.

\bibitem[Xu et~al.(2013)Xu, Jin, and Zhou]{Feature1}
Xu, M., Jin, R., and Zhou, Z.
\newblock Speedup matrix completion with side information: Application to
  multi-label learning.
\newblock In \emph{NeurIPS}, pp.\  2301--2309, 2013.

\bibitem[Ying et~al.(2018)Ying, He, Chen, Eksombatchai, Hamilton, and
  Leskovec]{PinSage}
Ying, R., He, R., Chen, K., Eksombatchai, P., Hamilton, W.~L., and Leskovec, J.
\newblock Graph convolutional neural networks for web-scale recommender
  systems.
\newblock In Guo, Y. and Farooq, F. (eds.), \emph{SIGKDD}, pp.\  974--983.
  {ACM}, 2018.

\bibitem[Zhang \& Chen(2020)Zhang and Chen]{IGMC}
Zhang, M. and Chen, Y.
\newblock Inductive matrix completion based on graph neural networks.
\newblock In \emph{ICLR}, 2020.

\bibitem[Zheng et~al.(2016)Zheng, Tang, Ding, and Zhou]{MFAutoReg}
Zheng, Y., Tang, B., Ding, W., and Zhou, H.
\newblock A neural autoregressive approach to collaborative filtering.
\newblock In \emph{ICML}, pp.\  764--773, 2016.

\bibitem[Zhong et~al.(2018)Zhong, Song, Jain, and Dhillon]{NIMC}
Zhong, K., Song, Z., Jain, P., and Dhillon, I.~S.
\newblock Nonlinear inductive matrix completion based on one-layer neural
  networks.
\newblock \emph{CoRR}, abs/1805.10477, 2018.

\end{thebibliography}
\bibliographystyle{icml2021}
}

\clearpage
\appendix

\section*{Appendix}
\section{Discussion on Representation Power and Expressiveness}\label{appendix-express}

We compare IDCF with other related models from two perspectives in order to shed more lights on the advantages and differences of our model.

\begin{figure*}[tb!]
\centering
\includegraphics[width=0.95\textwidth,angle=0]{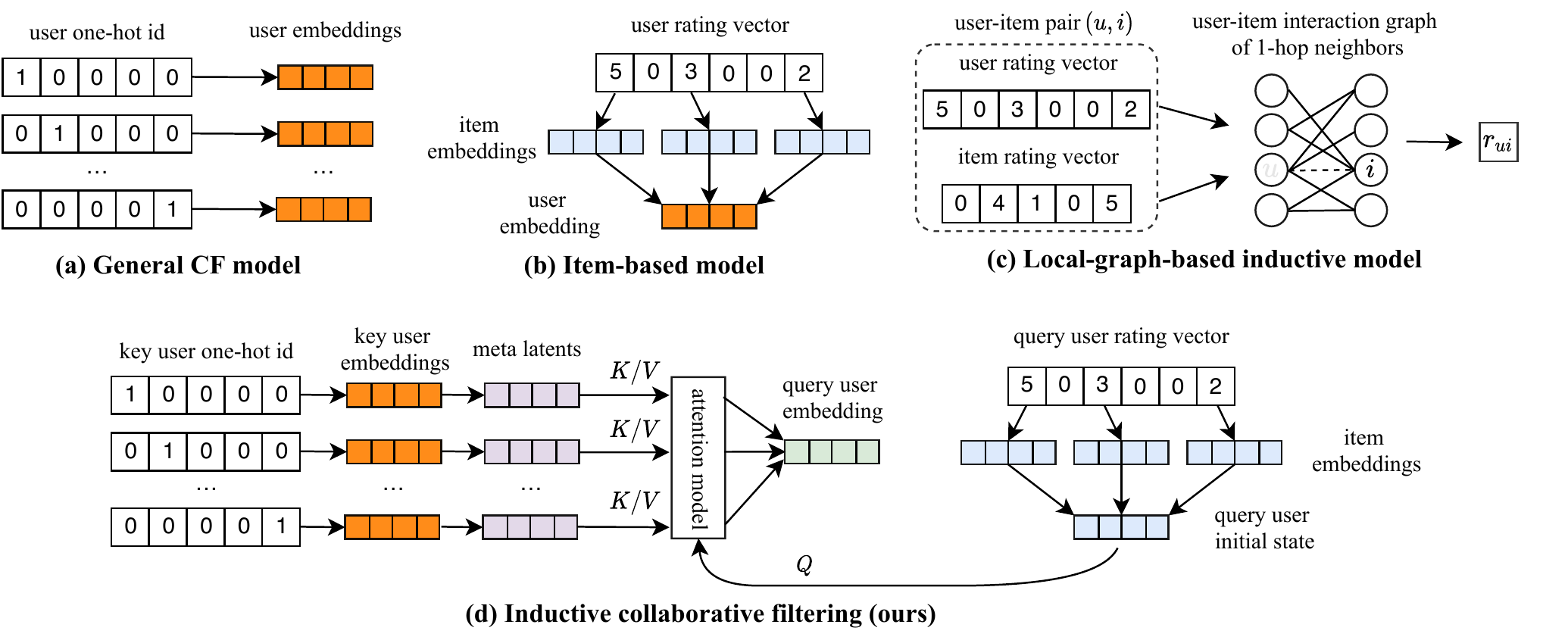}
\vspace{-10pt}
\caption{Comparison with related works on methodological level. We highlight the advantage of proposed model IDCF: it can 1) inductively compute user representations (or embeddings) and meanwhile 2) maintain enough capacity as general CF models.}
\label{fig-related}
\end{figure*}

\textbf{Comparison of Representation Power.}
We first provide a comparison with related works on methodological level as a clear elaboration for essential differences of our work to others. In Fig.~\ref{fig-related}, we present an intuitive comparison with general CF model, local-graph-based inductive CF and item-based CF model. As shown in Fig.~\ref{fig-related}(a), general collaborative filtering assumes user-specific embeddings for users and learn them collaboratively among all the users in one dataset. It disables inductive learning due to such learnable embeddings. Item-based model, as shown in Fig.~\ref{fig-related}(b), leverages embeddings of user's historically rated items to compute user's embeddings via some pooling methods. The learnable parameters only lie in the item space. It suffers from limited representation capacity compared to general CF model that assumes both user and item embeddings. Besides, local-graph-based inductive model (e.g. \cite{IGMC}), shown in Fig.~\ref{fig-related}(b), extracts local subgraph structures within 1-hop neighbors of each user-item pair (i.e., rated items of the user and users who rated the item) from a bipartite graph of all the observed user-item ratings and use GNNs to encode such graph structures for rating prediction. Note that the model requires that the local subgraphs do not contain user and item indices, so it cannot output user-specific embeddings. Its representation power is limited since it cannot represent diverse user preferences with arbitrary rating history on items. Also it cannot output user representations. Differently, our model IDCF, as shown in Fig.~\ref{fig-related}(d) adopts item-based embedding as initial states for query users to compute attention scores on key users and aggregate the embeddings (i.e., meta latents) of key users to estimate user-specific embeddings for query users, which maintains ability to produce user representations with enough representation power and meanwhile achieves inductive learning.

\begin{figure}[h]
\centering
\subfigure[]{
\begin{minipage}[t]{0.4\linewidth}
\centering
\includegraphics[width=0.95\textwidth,angle=0]{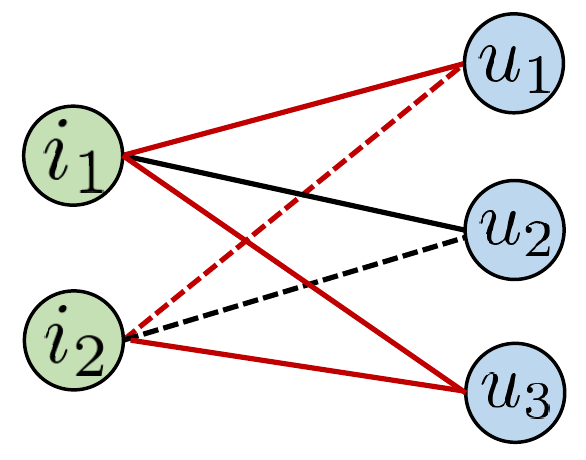}
\end{minipage}%
}%
\subfigure[]{
\begin{minipage}[t]{0.4\linewidth}
\centering
\includegraphics[width=0.95\textwidth,angle=0]{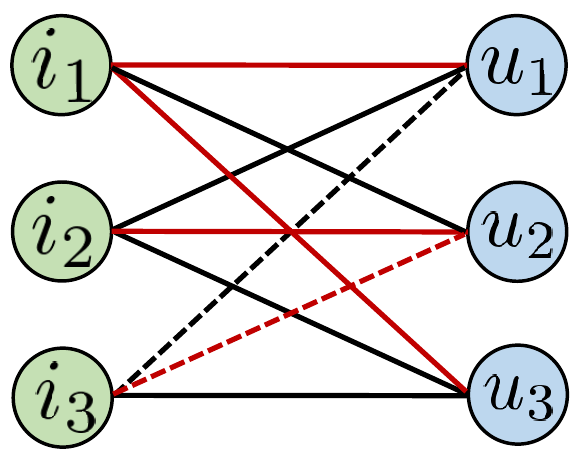}
\end{minipage}
}%
\vspace{-10pt}
\caption{Comparison of expressiveness. Feature-driven and local-graph-based models fail in (a) and (b), respectively. IDCF works effectively in both cases with superior expressiveness.}
\label{fig-express}
\end{figure}

\textbf{Comparison of Expressiveness.} We provide a comparison with feature-driven and local-graph-based inductive (matrix completion) models through two cases in Fig.~\ref{fig-express} so as to highlight the superior expressiveness of IDCF. Here we assume ratings are within $\{-1, 1\}$ (positive, denoted by red line, and negative, denoted by black line). The solid lines are observed ratings for training and dash lines are test ratings. In Fig.~\ref{fig-express}(a), we consider test ratings $(u_1, i_2)$ and $(u_2, i_2)$. For local-graph-based models, the 1-hop local subgraphs of $(u_1, i_2)$ (resp. $(u_2, i_2)$) consists of $\{i_1, i_2, u_1, u_3\}$ (resp. $\{i_1, i_2, u_2, u_3\}$) and the subgraph structures are different for two cases due to the positive (resp. negative) edge for $(u_1, i_1)$ (resp. $(u_2, i_1)$). The local-graph-based models can give right prediction for two ratings relying on different structures. Also, CF models and IDCF can work smoothly in this case, relying on different rating history of $u_1$ and $u_2$. However, feature-driven models will fail once $u_1$ and $u_2$ have the same features though two users have different rating history. In Fig.~\ref{fig-express}(b), we consider test ratings $(u_1, i_3)$ and $(u_2, i_3)$. The 1-hop local subgraphs of $(u_1, i_3)$ (resp. $(u_2, i_3)$) consists of $(i_1, i_2, i_3, u_1, u_3)$ (resp. $(i_1, i_2, i_3, u_2, u_3)$) and the subgraph structures are the same. Thus, the local-graph-based models will fail to distinguish two inputs and give the same prediction for $(u_1, i_3)$ and $(u_2, i_3)$. Differently, CF models and IDCF can recognize that $u_3$ has similar rating patterns with $u_1$ and different from $u_2$, thus pushing the embedding of $u_1$ (resp. $u_2$) close to (resp. distant from) $u_3$, which guides the model to right prediction. Note that the first case becomes a common issue when the feature space is small while the second case becomes general when the rating patterns of users are not distinct enough throughout a dataset, which induces similar local subgraph structures. Therefore, IDCF enjoys superior expressiveness for input data than feature-driven and local-graph-based inductive (matrix completion) models. Also, it maintains as good expressiveness as (transductive) CF models.

\section{Proofs in Section~\ref{sec-model}}\label{appendix-proof}

\subsection{Proof of Theorem 1}

\begin{proof}
The proof is trivial by construction. Assume the optimal $\mathbf P_2$ for Eq.~(\ref{eqn-obj3}) as $\mathbf P_2^*$. Since $\mathbf P_1$ given by Eq.~(\ref{eqn-obj1}) is column-full-rank, for any column vector $\mathbf p_{u'}^*$ in $\mathbf P_2^*$
($u'\in \mathcal U_2$), there exists $\mathbf c_{u'}^*$ such that $\mathbf c_{u'}^*{^\top}\mathbf P_1 = \mathbf p_{u'}^*$.
Hence, $\mathbf C^*=[\mathbf c_{u'}^*]_{u'\in \mathcal U_2}$ is a solution for Eq.~(\ref{eqn-obj2}) and gives $\mathcal D_{\mathcal S_q}(\hat R_2, R_2)<\epsilon$.
\end{proof}

\subsection{Proof of Theorem 2}

\begin{proof}

With fixed a true rating matrix $R_2$ to be learned and a probability distribution $\mathcal P$ over $[M_q]\times [N]$, which is unknown to the learner, we consider the problem under the framework of standard PAC learning. We can treat the matrix $R_2$ as a function $(u', i)\rightarrow r_{u'i}$. Let $\mathcal R$, a set of matrices in $\mathbb R^{M_q\times N}$, denotes the hypothesis class of this problem.  Then the input to the learner is a sample of $R_2$ denoted as
$$
\mathcal T = \left( (u'_t, i_t, r_{u'_ti_t}) | (u'_t, i_t)\in\mathcal S_q\right),
$$
where $\mathcal S_q=\{(u'_t,i_t)\}\in ([M_q]\times [N])^{T_2}$ is a set with size $T_2$ containing indices of the observed entries in $R_2$ and each $(u',i)$ in $\mathcal S_q$ is independently chosen according to the distribution $\mathcal P$. When using $\mathcal T$ as training examples for the learner, it minimizes the error $\mathcal D_{\mathcal S_q}(\hat R_2, R_2)=\frac{1}{T_2}\sum_{(u',i)\in \mathcal S_q}l(r_{u'i},\hat r_{u'i})$. We are interested in the generalization error of the learner, which is defined as
$$
\mathcal D(\hat R_2, R_2) = \mathbb E_{(u',i)\in \mathcal P} [l(r_{u'i}, \hat r_{u'i})].
$$

The (empirical) Rademacher complexity of $\mathcal R$ w.r.t. the sample $\mathcal T$ is defined as
$$
Rad_{\mathcal T}(\mathcal R) = \frac{1}{T_2} \mathbb E_{\sigma} \left[ \sup_{\hat R_2\in \mathcal R}\sum_{t=1}^{T_2} \sigma_t\hat r_{u'_ti_t}\right],
$$
where $\sigma_t\in\{-1, 1\}$ is a random variable with probability $Pr(\sigma_t=1)=Pr(\sigma_t=-1)=\frac{1}{2}$. Assume $l(\cdot, \cdot)$ is $L$-Lipschitz w.r.t. the first argument and $|l(\cdot,\cdot)|$ is bounded by a constant. Then a general result for generalization bound of $\mathcal R$ is
\paragraph{Lemma 1.}
\emph{
	 (Generalization bound \cite{foundation}): For a sample $\mathcal T$ with random choice of $\mathcal S_q=([M_q]\times [N])^{T_2}$, it holds that for any $\hat R_2\in \mathcal R$ and confidence parameter $0<\delta<1$,
	 \begin{equation}\label{eqn-lemma1}
	     Pr(\mathcal D(\hat R_2, R_2)\leq \mathcal D_{\mathcal S_q}(\hat R_2, R_2)+G)\geq 1 - \delta,
	 \end{equation}
	 where,
	 $$
	 G = 2L \cdot Rad(\mathcal R) + O\left(\sqrt{\frac{\ln(1/\delta)}{T_2}}\right).
	 $$
}

Based on the lemma, we need to further estimate the Rademacher complexity in our model to complete the proof. In our model, $\hat R_2 = \mathbf C^\top\mathbf P_k \mathbf Q$ and the entry $\hat r_{u'i}$ is given by $\hat r_{u'i}=\mathbf p_{u'}^\top \mathbf q_i = \mathbf c_{u'}^\top \mathbf P_k\mathbf q_i$ (where $\mathbf c_{u'}$ is the $u'$-th colunm vector of $\mathbf C$). Define $\mathcal C$ as a set of matrices,
$$
\mathcal C = \{\mathbf A\in [0,1]^{M_q\times M_k} :\|\mathbf a_{u'}\|_1 = \sum_{u=1}^{M_k} |a_{u'u}|=1\}.
$$
Then we have
\begin{equation}
\begin{aligned}
    T_2\cdot Rad_{\mathcal T}(\mathcal R) &= \mathbb E_\sigma \left[ \sup\limits_{\mathbf C\in\mathcal C}\sum\limits_{t=1}^{T_2} \sigma_t \mathbf c_{u'_t}^\top \mathbf P_k\mathbf q_{i_t} \right] \\
    & = \mathbb E_\sigma \left[ \sup\limits_{\mathbf C\in\mathcal C}\sum\limits_{u'=1}^{M_q} \mathbf c_{u'}^\top \cdot \left( \sum\limits_{t:u_t=u'}  \sigma_t  \hat R_{k, *i_t}\right) \right]\\  &(\mbox{since}\; \hat R_{k, *i_t}=\mathbf P_k\mathbf q_{i_t}) \\
    & \leq H\cdot \mathbb E_\sigma \left[ \sup\limits_{\mathbf A\in\mathcal A}\sum\limits_{u'=1}^{M_q} \mathbf a_{u'}^\top \cdot \left( \sum\limits_{t:u_t=u'}  \sigma_t  \hat R_{k, *i_t}\right) \right] \\
    & = H\cdot \mathbb E_\sigma \left[ \sum\limits_{u'=1}^{M_q} \max_{u\in [M_k]} \left( \sum\limits_{t:u_t=u'}  \sigma_t  \hat r_{ui_t}\right) \right].
\end{aligned}
\end{equation}
The last equation is due to the fact that $\mathbf a_{u'}$ is a probability distribution for choosing entries in $R_{k, *i_t}$, the $i_t$-th column of matrix $\hat R_k$. In fact, we can treat the $\max_{u\in [M_k]}$ inside the
sum over all $u'\in \mathcal U_2$ as a mapping $\kappa$ from $u'\in[M_q]$ to $u\in[M_k]$. Let $\mathcal K = \{\kappa: [M_q]\rightarrow [M_k]\}$ be the set of all mappings from $[M_q]$ to $[M_k]$, and then the above formula can be written as
\begin{align}
    &\mathbb E_\sigma \left[ \sum\limits_{u'=1}^{M_q} \max_{u\in [M_k]} \left( \sum\limits_{t:u_t=u'}  \sigma_t  \hat r_{ui_t}\right) \right]\\
    =&\mathbb E_\sigma \left[ \sup\limits_{\kappa\in \mathcal K} \sum\limits_{u'=1}^{M_q} \sum\limits_{t:u_t=u'}  \sigma_t  \hat r_{\kappa(u'),i_t}\right]\\
     =&\mathbb E_\sigma \left[ \sup\limits_{\kappa\in \mathcal K} \sum\limits_{t=1}^{T_2} \sigma_t \hat r_{\kappa(u_t),i_t}\right]\\
    \leq & B \sqrt{T_2} \cdot \sqrt{2M_q\log M_k}.
\end{align}
The last inequality is according to the Massart Lemma. Hence, we have
\begin{equation}\label{eqn-thm2-1}
    Rad_{\mathcal T}(\mathcal R) \leq H B\sqrt{\frac{2M_q\log M_k}{T_2}}.
\end{equation}
Incorporating Eq.~(\ref{eqn-thm2-1}) into Eq.~(\ref{eqn-lemma1}), we will arrive at the result in this theorem.
\end{proof}


\section{Extensions of IDCF}\label{appendix-extension}

IDCF can be extended to feature-based setting and deal with extreme cold-start recommendation where test users have no historical ratings. Here, we provide details of feature-based IDCF (IDCF-HY) which indeed is a hybrid model that considers both user features and one-hot user indices. Furthermore, we discuss in the views of transfer-learning and meta-learning that can be leveraged to enhance our framework as future study.

\subsection{Hybrid Model with Features (IDCF-HY)}\label{appendix-feature}

Assume $\mathbf a_u$ denotes user $u$'s raw feature vector, i.e., a concatenation of all the features (often including binary, categorical and continuous variables) where categorical features can be denoted by one-hot or multi-hot vectors. If one has $m$ user features in total, then $\mathbf a_u$ can be
$$
\mathbf a_u = [\mathbf a_{u1} || {\mathbf a_{u2}} || {\mathbf a_{u3}} || \cdots || {\mathbf a_{um}}].
$$
Then we consider user-sharing embedding function $\mathbf y_i()$ which can embed each feature vector into a $d$-dimensional embedding vector:
$$
\mathbf y_u = [\mathbf y_1(\mathbf a_{u1}) || \mathbf y_2({\mathbf a_{u2}}) || \mathbf y_3({\mathbf a_{u3}}) || \cdots || \mathbf y_m({\mathbf a_{um}})].
$$
Similarly, for item feature $\mathbf b_i = [\mathbf b_{i1} || {\mathbf b_{i2}} || {\mathbf b_{i3}} || \cdots || {\mathbf b_{in}}]$, we have its embedding representation:
$$
\mathbf z_i = [\mathbf z_1(\mathbf b_{i1}) || \mathbf z_2({\mathbf b_{i2}}) || \mathbf z_3({\mathbf b_{i3}}) || \cdots || \mathbf z_n({\mathbf b_{in}})].
$$
Also, we assume user-specific index embedding $\mathbf p_u$ and item-specific index embedding $\mathbf q_i$ for user $u$ and item $i$, respectively, as is in Section~\ref{sec-model}. The prediction for user $u$'s rating on item $i$ can be
\begin{equation}
    \hat r_{ui} = g_\theta(\mathbf p_u, \mathbf y_u, \mathbf q_i, \mathbf z_i),
\end{equation}
where $g_\theta$ can be a shallow neural network with parameters denoted by $\theta$. To keep notation clean, we denote $\mathbf Y=\{\mathbf y_1, \mathbf y_2, \cdots, \mathbf y_m\}$ and $\mathbf Z=\{\mathbf z_1, \mathbf z_2, \cdots, \mathbf z_n\}$. Then for key users in $\mathcal U_k$ with rating matrix $R_k$, we consider the optimization problem,
\begin{equation}\label{eqn-feature-opt1}
    \min_{\mathbf P_k, \mathbf Q, \mathbf Y, \mathbf Z, \theta} \mathcal D_{\mathcal S_k}(\hat R_k, R_k),
\end{equation}
based on which we get learned feature embedding functions $\mathbf Y$, $\mathbf Z$ as well as transductive embedding matrices $\mathbf P_k$, $\mathbf Q$ which we further use to compute inductive embeddings for query users.

For query users, feature embeddings can be obtained by the learned $\mathbf Y$ and $\mathbf Z$ in Eq.~(\ref{eqn-feature-opt1}), i.e., $\mathbf y_{u'}=[\mathbf y_{u'1}(\mathbf a_{u'1}) || \cdots || \mathbf y_{u'm}(\mathbf a_{u'm})]$ where $\mathbf a_{u'}$ is raw feature vector of user $u'$. Then we have a relation learning model $h_w$ that consists of a multi-head attention function and use user feature as input $\mathbf d_{u'}=\mathbf y_{u'}$. The inductive user-specific representation can be given by $\mathbf p_{u'}=h_w(\mathbf d_{u'})$ (i.e., Eq.~(\ref{eqn-attn2}) and Eq.~(\ref{eqn-induc_embed})), similar as the CF setting in Section~\ref{sec-model}. The rating of user $u'$ on item $i$ can be predicted by $\hat r_{u'i}=g_\theta(\mathbf p_{u'}, \mathbf y_{u'}, \mathbf q_i, \mathbf z_i)$. Also, the optimization for inductive relation model is
\begin{equation}
    \min_{w, \theta} \mathcal D_{\mathcal S_q} (\hat R_q, R_q).
\end{equation}

\subsection{Extreme Cold-Start Recommendation}\label{appendix-zero}

For cold-start recommendation where test users have no historical rating, we have no information about users if without any side information. In such case, most CF models would fail for personalized recommendation and degrade to a trivial one which outputs the same result (or the same distribution) to all the users using the popularity of items. For IDCF, the set $\mathcal I_{u'}$ in Eq.~(\ref{eqn-attn}) would be empty for users with no historical rating, in which situation we can randomly select a group of key users to construct $\mathcal I_{u'}$ used for computing attentive scores with key users. Another method is to directly use average embeddings of all the key users as estimated embeddings for query users. In such case, the model degrades to ItemPop (using the numbers of users who rated the item for prediction).

On the other hand, if side information (such as user profile features) is available, our hybrid model IDCF-HY can leverage user features for computing inductive embeddings, which enables extreme cold-start recommendation. We apply this method to cold-start recommendation on Movielens-1M using features and the results are given in Appendix~\ref{appendix-moreres}. 

\subsection{Transfer Learning \& Meta-Learning}\label{appendix-meta}

Another extension of IDCF is to consider transfer learning on cross-domain recommendation tasks or when treating recommendation for different users as different tasks like \cite{MeLU}. Transfer learning and meta learning have shown power in learning generalizable models that can adapt to new tasks. In our framework, we can also take advantage of transfer learning (few-shot learning or zero-shot learning) or mete-learning algorithms to train our relation learning model $h_w$. For example, if using model-agnostic meta-learning algorithm for the second-stage optimization, we can first compute one-step (or multi-step) gradient update independently for each user (or a group of clustering users) in a batch and then average them as one global update for the model. The meta-learning can be applied over different groups of users or cross-domain datasets.

\section{Details in Implementations}\label{appendix-implement}

We provide implementation details that are not presented in Section~\ref{sec-exp} in order for reproducibility.

\subsection{Hyper-parameter Settings}\label{appendix-paras}

We present details for hyper-parameter settings in different datasets. We use $L=4$ attention heads for our inductive relation learning model among all the datasets. For Douban and ML-100K, each attention head randomly samples 200 key users for computing attention weights. For ML-1M, we set sample size as 500; for Amazon-Books and Amazon-Beauty, we set it as 2000. We use Adam optimizer and learning rates are searched within [0.1, 0.01, 0.001, 0.0001]. For pretraining, we consdier L2 regularization for user and item embeddings. The regularization weights are searched within [0.001, 0.002, 0.005, 0.01, 0.02, 0.05, 0.1, 0.2]. The mini-batch sizes are searched within [64, 256, 512, 1024, 2048] to keep a proper balance between training efficiency and performance. For adaption stage, regularization weight $\lambda$ is set as 10 for five datasets. Besides, different hyper-parameters for architectures are used in three implementations.   

\textbf{IDCF-NN.}
For Douban and ML-100K, we use embedding dimension $d=16$ and neural size $48-32-32-1$ for $f_\theta$. For ML-1M, ML-10M and Amazon-Books, we use $d=32$ and neural size $96-64-64-1$ for $f_\theta$.  

\textbf{IDCF-GC.}
For Douban and ML-100K, we use embedding dimension $d=32$ and neural size $128-32-32-1$ for $f_\theta$. For ML-1M, ML-10M and Amazon-Books, we use $d=64$ and neural size $256-64-64-1$ for $f_\theta$. 

\textbf{IDCF-HY.}
We use embedding size $d=32$ for each feature in ML-1M as well as user-specific and item-specific index embeddings. The neural size of $g_\theta$ is set as $320-64-64-1$.

\subsection{Evaluation Metrics}\label{appendix-metric}
We provide details for our adopted evaluation metrics. In our experiments, we follow evaluation protocols commonly used in previous works in different settings. Three metrics used in our paper are as follows.

\begin{itemize}
\item \textbf{RMSE:} Root Mean Square Error is a commonly used metric for explicit feedback data and measures the averaged L2 distance between predicted ratings and ground-truth ratings:
\begin{equation}
     RMSE = \sqrt{\frac{\sum\limits_{(u,i)\in \mathcal{I}^+}(\hat r_{ui} - r_{ui})^2}{|\mathcal{I}^+|}}.
\end{equation}

\item \textbf{AUC:} Area Under the ROC Curve is a metric for implicit feedback data. It measures general consistency between a ranking list of predicted scores and ground-truth ranking with 1's before 0's. More specifically, AUC counts the average area under the curve of true-positive v.s. false-positive curve:
		\begin{equation}
		AUC = \frac{\sum_{(u,i) \in \mathcal{I}^+} \sum_{(u',i') \in \mathcal{I}^-} \delta(\hat{r}_{u,i} > \hat r_{u',i'})}{|\mathcal{I}^+||\mathcal{I}^-|},
		\end{equation}
		where $\mathcal{I}^+ = \{(u,i)|r_{ui} > 0\}$ and $\mathcal{I}^- = \{(u',i')|r_{u'j'} = 0\}$ denote the sets of observed user-item interaction pairs and unobserved user-item pairs respectively. The indicator $\delta(\hat{r}_{ui} > \hat{r}_{uj})$ returns 1 when $\hat{r}_{ui} > \hat{r}_{uj}$ and 0 otherwise. 
Since we only have ground-truth positive examples (clicked items) for users, we negatively sample five items as negative examples (non-clicked items) for each user-item rating in dataset, which composes the set $\mathcal I^-$.

\item \textbf{NDCG:} Normalized discounted cumulative gain (Normalized DCG) measures the usefulness, or gain, of a recommended item based on its position in the result list. NDCG can be used to evaluate the model on both explicit and implicit feedback data. The gain is accumulated from the top of the recommendation list to the bottom, with the gain of each result discounted at lower ranks. The NDCG metric is computed for each user and can measures the averaged performance of personalized recommendation. Given the ranking list of recommended items to a user $u$, denoted as $\hat {\mathcal K}_u$, its $DCG$ is defined as:
\begin{equation}
     DCG = \sum\limits_{i\in \hat{\mathcal K}_u}\frac{rel_i}{\log_2(i+1)}.
\end{equation}
 where $rel_i$ is the graded relevance of item $i$ with user $u$. For explicit feedbacks, $rel_i$ is the ground-truth rating of user $u$ on item $i$. For implicit feedbacks, $rel_i=1$ for observed user-item interaction and $rel_i=0$ otherwise. The normalized discounted cumulative gain, or NDCG, is computed as:
\begin{equation}
     NDCG@K = \frac{DCG@K}{IDCG@K}
\end{equation}
where $IDCG@K$ is ideal discounted cumulative gain:  $IDCG@K = \sum\limits_{i\in \mathcal K_u}\frac{rel_i}{\log_2(i+1)}$, and $\mathcal K_u$ represents the ground-truth ranking list of relevant items (ordered by their ground-truth ratings/interactions by user $u$).


\end{itemize}
\begin{table*}[t!]    
\centering
\small
\caption{\label{tab-split} Test RMSEs on all the users (All), few-shot query users (FS) and new users (New) of IDCF-NN in Movielens-1M using different splits for key and query users. (Lower RMSE is better)}   
\begin{tabular}{c|c|c|c|c|c|c|c}
\cline{1-8}
\multirow{4}{*}{Threshold} & $\delta$ & 20 & 30 & 40 & 50 & 60 & 70 \\
\cline{2-8}
& All (RMSE)         & 0.8440 & 0.8437 & 0.8439 & 0.8440 & 0.8444 & 0.8451 \\
\cline{2-8}
& FS (RMSE)       & 0.9785 & 0.9525 & 0.9213 & 0.9166 & 0.9202 & 0.9160 \\
\cline{2-8}
& New (RMSE)         &  0.9945 & 0.9942 & 0.9902 & 0.9883 & 0.9911 & 0.9929\\
\cline{1-8}
\multirow{4}{*}{Random} & $\gamma$ & 0.97 & 0.85 & 0.75 & 0.68 & 0.62 & 0.57 \\
\cline{2-8}
& All (RMSE)         & 0.8446 & 0.8536 & 0.8587 & 0.8637 & 0.8669 & 0.8689 \\
\cline{2-8}
& FS (RMSE)       & 0.8863 & 0.8848 & 0.8760 & 0.8805 & 0.8824 & 0.8855 \\
\cline{2-8}
& New (RMSE)         &  0.9901 & 0.9923 & 0.9956 & 0.1001 & 1.0198 & 1.0262  \\
\cline{1-8}

\end{tabular}  
\end{table*}


\section{More Experiment Results}\label{appendix-moreres}
\subsection{Impacts of different splits for key and query users}
In our experiments in Section~\ref{sec-exp}, we basically consider users with more than $\delta$ training ratings as $\mathcal U_1$ and the remaining as $\mathcal U_2$, based on which we construct key users and query users to study model's performance on few-shot query users (for inductive interpolation) and new unseen users (for inductive extrapolation). Here we provide a further discussions on two spliting ways and study the impact on model performance.
\begin{itemize}
    \item \textbf{Threshold}: we select users with more than $\delta$ training ratings as $\mathcal U_1$ and users with less than  $\delta$ training ratings as $\mathcal U_2$.
    \item \textbf{Random}: we set a ratio $\gamma\in (0,1)$ and randomly sample $\gamma \times 100\%$ of users in the dataset as $\mathcal U_1$. The remaining users are grouped as $\mathcal U_2$.
\end{itemize}
We consider $\delta=[20, 30, 40, 50, 60, 70]$ and $\gamma=[0.97, 0.85, 0.75, 0.68, 0.62, 0.57]$ (which exactly gives the same ratio of $|\mathcal U_1|$ and $|\mathcal U_2|$ as corresponding $\delta$ in threshold split\footnote{For example, using $\gamma=0.97$ in random split will result in the same sizes of $\mathcal U_1$ and $\mathcal U_2$ as using $\delta=20$ in threshold split.}) in Movielens-1M dataset. For each spliting, we also consider two situations for key users and query users: 1) inductive learning for interpolation on few-shot query users, i.e., the first-starge training is on the training ratings of key users $\mathcal U_k = \mathcal U_1$ and the second-stage training is on the training ratings of query users $\mathcal U_k = \mathcal U_2$; inductive learning for extrapolation on zero-shot new users, i.e., the two-stage trainings are both on the training ratings of a same group of users $\mathcal U_k = \mathcal U_q = \mathcal U_1$. We test the model on the testing ratings of users in $\mathcal U_2$. The results of IDCF-NN are presented in Table \ref{tab-split} where we report test RMSEs on all the users, few-shot query users and zero-shot new users.

As we can see from Table \ref{tab-split}, with \emph{threshold split}, as $\delta$ increases (we have fewer key users and more query users and they both have more training ratings on average), test RMSEs for query users exhibit a decrease. The reason is two-folds: 1) since key usres have more training ratings, the transductive model can learn better representations; 2) since query users have more training ratings, the inductive model would have better generalization ability. On the other hand, with different spliting thresholds, test RMSEs for new users remain in a fixed level. The results demonstrate that the performance of IDCF on new unseen users is not sensitive to different splitting thresholds. However, with \emph{random split}, when $\gamma$ decreases (also we have fewer key users and more query users but their average training ratings stay unchanged), RMSEs for new users suffer from an obvious decrease. One possible reason is that when we use smaller ratio of key users with random split, the `informative' key users in the dataset are more likely to be ignored. (Recall that, as is shown in Fig.~\ref{fig-attn-a}, there exist some important key users that give high attention weights on query users.) If such key users are missing, the performance would be affected due to insufficient expressive power of the inductive representation model.

Comparing threshold split with random split, we can find that when using the same ratio of key users and query users (i.e., the same column in Table \ref{tab-split}), RMSEs on new users with threshold split are always better than those with random split. Such observation again shows that key users with more historical ratings would be more informative for providing useful information to inductive representation learning on query users, and again echo the results in Fig.~\ref{fig-attn-b} which demonstrates that important key users who give large attention weights on query users tend to exist in users with sufficient historical ratings.

\subsection{Ablation Studies}

\begin{table*}
	\centering
	\caption{Ablation studies on ML-1M and Amazon-Books datasets. (Lower RMSE and Higher AUC/NDCG are better)\label{tbl-abl}}
        \begin{tabular}{c|cc|cc|cc|cc}
            \hline
            \multirow{3}{*}{\textbf{Method}}& \multicolumn{4}{c|}{\textbf{ML-1M}} & \multicolumn{4}{c}{\textbf{Amazon-Books}} \\
			\cline{2-9}
			& \multicolumn{2}{c|}{\textbf{Query}} & \multicolumn{2}{c|}{\textbf{New}} & \multicolumn{2}{c|}{\textbf{Query}} & \multicolumn{2}{c}{\textbf{New}} \\
			\cline{2-9}
			& \textbf{RMSE} & \textbf{NDCG} & \textbf{RMSE} & \textbf{NDCG} & \textbf{AUC} & \textbf{NDCG} & \textbf{AUC} & \textbf{NDCG} \\
			\hline
			\textbf{IDCF-GC} & \textbf{0.944} & \textbf{0.940} & \textbf{0.957} & \textbf{0.942} & \textbf{0.938} & \textbf{0.946} & \textbf{0.921} & \textbf{0.930} \\
			\hline
			\textbf{RD-Item} & 1.014 & 0.843 & 1.023 & 0.835 & 0.665 & 0.701 & 0.832 & 0.821 \\
			\textbf{Trans-User} & 0.992 & 0.876 & - & - & 0.845 & 0.821 & - & -  \\
			\textbf{Meta-Path} & 0.959 & 0.912 & 0.981 & 0.892 & 0.910 & 0.916 & 0.882 & 0.901  \\
			\hline
		\end{tabular}
\end{table*}

In Table~\ref{tbl-abl} we present the results of ablation study on ML-1M and Amazon-Books datasets. We compare IDCF-GC with 1) RD-Item (using randomized item embeddings), 2) Trans-User (directly optimizing $\mathbf P_q$ in Eqn.~(\ref{eqn-obj2})) and 3) Meta-Path (using meta-path \emph{user-item-user} in the observed user-item bipartite graph to determine users' neighbors for message passing). The results show that RD-Item performs much worse than IDCF-GC since the randomized item embeddings may provide wrong signals for both graph learning and final prediction. Compared with Trans-User, IDCF-GC significantly outperforms it over a large margin. The reason is that directly optimizing $\mathbf P_q$ would lead to serious over-fitting since query users have few training data. Furthermore, we can see that Meta-Path provides inferior performance than IDCF-GC in Table~1. The reason is that meta-path can only identify limited relations from observed bipartite graph that often has missing/noisy links, while IDCF learns and explores useful semantic relations for sufficient message passing.

\subsection{Cold-Start with User Features}
We also wonder if our inductive model can handle extreme cold-start users with no historical rating \footnote{In some literature, cold-start users also mean users with few historical ratings (for training or/and inference). Here we consider extreme cold-start recommendation for users with no historical rating for both training and inference.}. Note that cold-start users are different and more challenging compared to new (unseen) users. For new users, the model can still use historical ratings as input features during inference, though it cannot be trained on these ratings. To enable cold-start recommendation, we leverage user attribute features in Movielens-1M. We use the dataset provided by \cite{MeLU}, which contains attribute features and split warm-start and cold-start users. For IDCF, we also adopt the training algorithm of inductive learning for extrapolation and treat the warm-start users as key and query users. We user the \emph{warm-start} users' training ratings for model training and the \emph{cold-start} users' test ratings for test. We compare with Wide\&Deep network \cite{Feature2}, GCMC (using feature vectors) and two recently proposed methods for cold-start recommendation: graph-based model AGNN \cite{CSR-GNN} and meta-learning model MeLU \cite{MeLU}. 

\begin{figure}[t!]
\begin{minipage}{0.99\linewidth}
\centering
\subfigure[Test RMSEs]{
\begin{minipage}[t]{0.48\linewidth}
\centering
\label{fig-rmse}
\includegraphics[width=0.95\textwidth,angle=0]{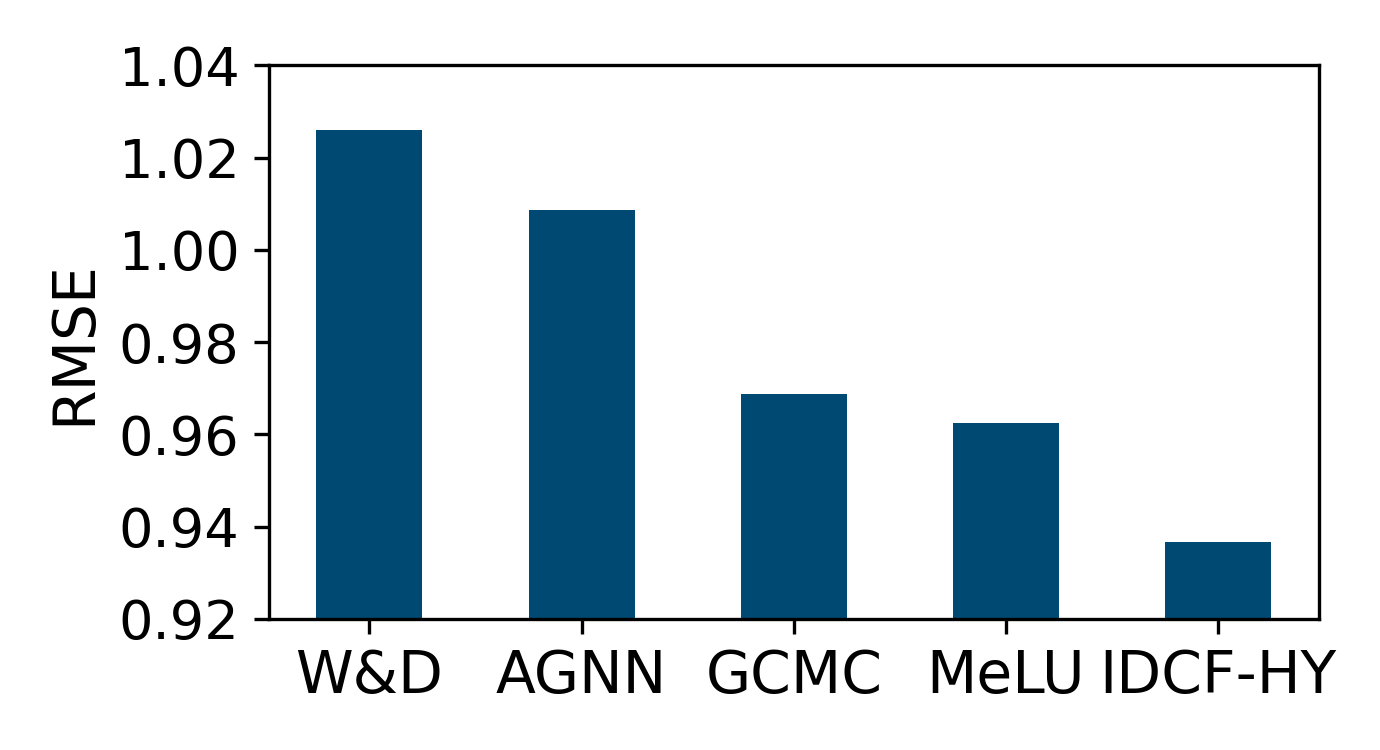}
\end{minipage}%
}%
\subfigure[Training time per epoch]{
\begin{minipage}[t]{0.48\linewidth}
\centering
\label{fig-time}
\includegraphics[width=0.95\textwidth,angle=0]{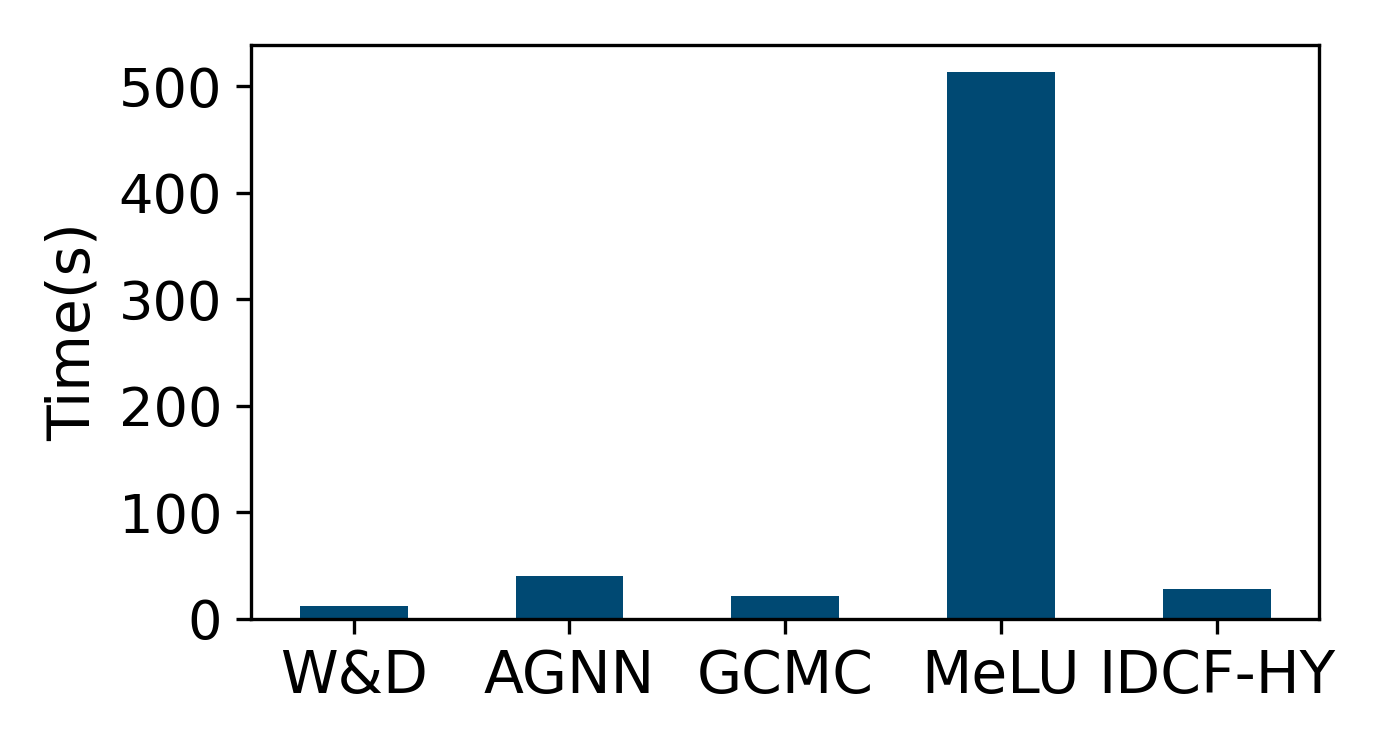}
\end{minipage}%
}%
\end{minipage}
\vspace{-10pt}
\caption{Performance comparison for extreme cold-start recommendation on ML-1M with user profile features.}
\vspace{-10pt}
\end{figure}

Fig.~\ref{fig-rmse} gives the test RMSEs for all the models. It shows that our IDCF-HY outperforms the competitors, achieving $2.6\%$ improvement of RMSE over the best one MeLU even on the difficult zero-shot recommendation task. The result indicates that IDCF is a promising approach to handle new users with no historical behavior in real-world dynamic systems. We also compare the training time per epoch of each method in Fig.~\ref{fig-time}. By contrast, IDCF-HY is much faster than MeLU and as efficient as AGNN and GCMC.

\end{document}